\documentclass[a4paper,12pt]{article}
\usepackage{amssymb,amsmath,amsfonts,amsthm,amscd,epsfig,bbold,stmaryrd,bm}
\usepackage{dsfont}
\usepackage{multirow}

\title{Spin dynamics in finite cyclic $XY$ model}

\author{Evgeniy Safonov\hspace*{1mm}$^{\rm a}$,
 Oleg Lychkovskiy
\hspace*{1mm}$^{\rm b,a}$
\\
${\rm ^a}$
{\small\it Institute for Theoretical and Experimental Physics}\\
{\small\it 117218, B.Cheremushkinskaya 25,
Moscow, Russia}
\\
${\rm ^b}$
{\small\it
Physics Department, Lancaster University, Lancaster, LA1 4YB, UK}
}

\date{}

\begin{document}
\maketitle

\renewcommand{\Re}{\mathrm {Re}}
\renewcommand{\Im}{\mathrm {Im}}

\newcommand{\be}{\begin{equation}}
\newcommand{\ee}{\end{equation}}

\newcommand{\ii}{\mathrm{i}}
\newcommand{\pp}{\mathbf{p}}
\newcommand{\ssigma}{\bm{\sigma}}
\newcommand{\q}{\mathbf{q}}
\newcommand{\e}{\mathbf{e}}
\newcommand{\A}{\mathcal{A}}

\newcommand{\re}{\mathrm{Re}}
\newcommand{\im}{\mathrm{Im}}

\newcommand{\p}{\tilde{p}}
\newcommand{\x}{\tilde{x}}

\newcommand{\la}{\langle}
\newcommand{\ra}{\rangle}

\newcommand{\cH}{{\cal H}}
\newcommand{\cS}{{\cal S}}
\newcommand{\cB}{{\cal B}}
\newcommand{\cF}{{\cal F}}

\newcommand{\hPS}{ P^{\cal S}}
\newcommand{\hPE}{ P^{\cal E}}
\newcommand{\hH}{ H}
\newcommand{\hHB}{ H^{\cal B}}
\newcommand{\hHS}{ H^{\cal S}}
\newcommand{\hHSB}{ H^{\cal SB}}
\newcommand{\hV}{ V}
\newcommand{\hS}{ S}
\newcommand{\psiS}{\psi^{\cal S}}
\newcommand{\psiB}{\psi^{\cal B}}

\newcommand{\dens}{ \rho}
\newcommand{\densS}{ \rho^{\cal S}}
\newcommand{\densB}{ \rho^{\cal B}}
\newcommand{\meandens}{{ \bar \rho}}
\newcommand{\meandensS}{\overline{{ \rho}^{\cal S}}}
\newcommand{\meanavdensS}{\langle \overline{{ \rho}^{\cal S}}\rangle_{\cal B}}

\newcommand{\tr}{\mathrm{tr}}
\newcommand{\trS}{\mathrm{tr}_{\cal S}}
\newcommand{\trE}{\mathrm{tr}_{\cal E}}
\newcommand{\Pu}{\mathfrak P}

\begin{abstract}
Evolution of the $z$-component of a single spin in the finite cyclic $XY$ spin $1/2$ chain is studied. Initially one selected spin is polarized while other spins are completely unpolarized and uncorrelated. Polarization of the selected spin as a function of time is proportional to the autocorrelation function $g^{zz}_0(t)$ at infinite temperature. Initialization of the selected spin gives rise to two wave packets moving in opposite directions and winding over the circle. We express $g^{zz}_0(t)$ as a series in winding number and derive tractable approximations for each term.
This allows to give qualitative explanation and quantitative description to various finite-size effects such as partial revivals and transition from regular to erratic behavior.
\end{abstract}


\section{Introduction}


Exactly solvable spin chains are widely used as toy models for exploring various aspects of quantum dynamics. Recent progress in experimental techniques allows to construct quantum systems with effective spin chain Hamiltonians~(see e.g. \cite{PorrasCirac}), which opens new prospects for exploring fundamental concepts such as decoherence and thermalization, as well as for applications such as quantum state transfer through quantum wires~\cite{bose2003quantum}.
This motivates further efforts to understand dynamics of spin chains in detail.

We consider the reduced dynamics of a single spin in the cyclic spin $1/2$ $XY$ chain with finite number of spins, $N.$ Initially one selected spin has a given polarization while other $(N-1)$ spins are completely uncorrelated and unpolarized. We study the $z$-component of polarization of a spin as a function of time. It can be expressed through the two-spin time-dependent correlation functions $g^{zz}_n(t).$ Although many papers starting from the  pioneering paper on $XY$ chain \cite{lieb1961two}  were devoted to calculation of  various correlation functions, most of the studies concentrated on the thermodynamic limit \mbox{$N\rightarrow\infty.$} Exact expression for $g^{zz}_n(t)$ in the $XY$ model with finite $N$  was derived in \cite{mazur1973time,siskens1974time}. It involves sums of $\sim N$ oscillating terms and thus is hardly tractable.
However, these sums can be calculated numerically for various values of model parameters. Resulting plots for  $g^{zz}_n(t)$ readily reveal a rich variety of spin evolution patterns which call for explanation (see figures in the present paper, especially fig. \ref{fig variety of patterns}). One striking feature of the evolution is regular-to-erratic transition: $g^{zz}_n(t)$ is described fairly well by $N\rightarrow\infty$ approximation (which is given by a rather regular function of time for a wide range of model parameters) up to some threshold time $t_{\rm th},$ but at $t_{\rm th}$ this concordance is abruptly destroyed by sharp revival;
at later times the evolution becomes less and less regular and ends up with apparently chaotic fluctuations near the long-time average. This feature is apparently common for all finite spin chains; in particular, it was observed in numerical simulations done for the $XX$ (isotropic $XY$) model \cite{fel1998regular, fel1999regular}, for the $XXZ$ model with long-range \cite{bruschweiler1997non} and nearest-neighbor \cite{Mossel} couplings, for the $XY$ model \cite{lychkovskiy2011entanglement}.

Results of the numerical studies and general considerations suggest that it is the winding of two oppositely directed wave packets created by the spin initialization which underlies the large-time dynamics in the cyclic chain  \cite{fel1999regular} (in case of an open-ended chain the same role is played by the reflection of the packets from the ends of the chain \cite{fel1998regular}). Threshold time corresponds to the time necessary for a forefront of a wave packet to make one round trip over the circle.\footnote{See also a recent paper \cite{happola2012universality} for the same physical reasoning applied to dynamics after a quench in the $XY$ model.} The interference between the forefronts of the  wave packets and their own tails produces partial revivals at $t=t_{\rm th},2t_{\rm th},...$ and leads to the regular-to-erratic transition.

In order to study spin dynamics in finite spin chains at times greater than $t_{\rm th}$ it is desirable to have tractable analytical approximations for $g^{zz}_n(t)$ valid for $t>t_{\rm th}.$ The main goal of the present paper is to obtain such approximations for the cyclic $XY$ model. The mathematical method which we use and develop is closely related to the physical picture of wave packet winding over the circle and in fact allows to quantitatively describe such winding. Namely, we are able to represent the correlation function as a series in winding number $s.$ This series has an appealing property that $(s+1)$ first terms are enough to describe the correlation function for $t<(s+1)t_{\rm th}.$  Such truncated series fully takes into account interference between the components of the wave packets which have completed  $0,1,2,...,s$ round trips over the circle. These approximations are  fairly accurate even when the evolution is already completely irregular.
A related result in this direction was previously obtained in a special case of $XX$ model \cite{Zhu2010XX, benderskii2011propagating}: a quasi-particle Green's function was represented as a sum over winding numbers. Recently the method was applied to the inhomogeneous open-ended $XX$ chain \cite{benderskii2011propagating,benderskii2013propagation}. Similar mathematical structures and physical patterns emerge in the systems of coupled oscillators, see e.g. ref. \cite{benderskii2013revivals} and references therein.

The approximation accounting for $s$ windings involves $\sim s$ oscillating terms and therefore is much more tractable than the exact formula as long as $s\ll N.$ This allows to look at the regular-to-erratic transition (as well as on some other peculiar features of spin evolution in finite chains) from a new perspective and obtain new quantitative results hardly accessible in numerical simulations. For example, we are able to derive an asymptotic formula for the amplitude of the $s$'th revival.

We also touch the issue of incomplete thermalization of spins in the $XY$ spin chain. In particular we show that the autocorrelation function $g^{zz}_0(t)$ at infinite temperature never changes its sign in contrast to what should be expected in case of complete thermalization. This intriguing property was previously proven in the special case of the $XX$ chain \cite{fel1998regular,fel1999regular} and observed in numerical calculations of spin evolution in the $XXZ$ model with long-range interactions~\cite{bruschweiler1997non}.

The rest of the paper is organized as follows. In Sec. \ref{sec XY model} we briefly describe the $XY$ model on a circle. In Sec. \ref{sec exact formulas} we discuss exact formula for $g^{zz}_n(t)$ and rewrite it through sums over winding numbers. In two special cases (the Ising model with critical magnetic field and the $XX$ model) this directly leads to the desired result: $g^{zz}_n(t)$ is represented in a transparent and convenient way through the infinite sum of Bessel functions; such representation allows to obtain simple successive approximations valid up to times $t_{\rm th},2t_{\rm th},...$. However in the general case each term of the sum is represented as an integral which should be worked out.
In Sec. \ref{sec Asymptotics} we handle these integrals approximately. Thus we obtain our main result -- the successive asymptotic approximations in a general case.
In Sec. \ref{sec Chaos} we discuss the transition from regular to erratic behavior.
The results are summarized in  Sec. \ref{sec Conclusions}. Bulk of technical details is presented in Appendices.  In Appendix \ref{appendix Diagonalization} we describe the diagonalization of the $XY$ model.  In Appendix \ref{appendix Exact correlation function} we rederive the exact formula for  $g^{zz}_n(t)$ at infinite temperature using a method which is somewhat more direct than one implemented in the original work \cite{mazur1973time,siskens1974time}. These two appendices mostly contain widely known calculations and results; however we include them in order to introduce our notations, to emphasize some salient features usually omitted in the literature and for the sake of completeness. In Appendix~\ref{appendix velocity} the dependence of the wave packet forefront velocity on the model parameters is investigated.  Appendix~\ref{appendix Asymptotics} contains the technical details of calculating asymptotic expressions presented in Sec. \ref{sec Asymptotics}.

\section{\label{sec XY model}$XY$ model on a circle}

We consider a chain of $N$ coupled spins $1/2$ with the following Hamiltonian \cite{lieb1961two,niemeijer1967some}:
\be
\hH =\frac14  \sum_{n=1}^N ((1+\gamma)\sigma_n^x \sigma_{n+1}^x+ (1-\gamma) \sigma_n^y \sigma_{n+1}^y)  + \frac{h}2  \sum_{n=1}^N \sigma_n^z.
\ee
Here the index $N+1$ is identified with $1,$ and $N$ is supposed to be even. Two parameters enter the Hamiltonian, the anisotropy parameter $\gamma$ and the magnetic field $h.$ Without loss of generality one may assume $\gamma\geq0$, $h\geq0$ (see Appendix~\ref{appendix Diagonalization}). In Sec. \ref{sec Asymptotics} we will concentrate on the case $h\geq1,$ $\gamma\in[0,1].$


An important property of the $XY$ Hamiltonian is that it commutes with the parity operator $\Pi\equiv \prod_{n=1}^N \sigma^z.$
It can be represented in an "almost free-fermion form" through the sequential Jordan-Wigner, Fourier and Bogolyubov transformations \cite{lieb1961two,mazur1973time,siskens1974time} (see appendix~\ref{appendix Diagonalization} for the details):
%
\be
H=P^{\rm odd} \sum_{q\in X_{\rm odd}} E_q (c_{q}^+ c_{q}-\frac12) + P^{\rm ev} \sum_{q\in X_{\rm ev}} E_q (c_{q}^+ c_{q}-\frac12),
\ee
where
\be
X_{\rm odd}=\{-\frac{N}2+1,-\frac{N}2+2,...,\frac{N}2\},
~~~X_{\rm ev}=\{-\frac{N}2+\frac12,-\frac{N}2+\frac32,...,\frac{N}2-\frac12\},
\ee
$\{ c_{q},~ q\in X_{\rm odd} \}$ and $\{ c_{q},~q\in X_{\rm ev} \}$ are two sets of fermion operators (note, however, that two operators from different sets do not satisfy fermion anticommutation relations, see eq.(\ref{c anticommutation})), $P^{\rm odd}$ and $P^{\rm ev}$ are parity projectors,
\be
P^{\rm ev}\equiv (1+\Pi)/2,~~~~ P^{\rm odd}\equiv (1-\Pi)/2,
\ee
and fermion energy is defined as $E_q\equiv E(\varphi(q)),$ $\varphi(q)\equiv\frac{2\pi q}{N},$
\be
 E(\varphi)=\sqrt{\varepsilon(\varphi)^2+\Gamma(\varphi)^2},~~\varepsilon(\varphi)\equiv h-\cos\varphi,~~\Gamma(\varphi)\equiv \gamma \sin\varphi.
\ee
One can see that the Hilbert space is divided into a two subspaces with odd and even numbers of fermions correspondingly. Number of fermions is an integral of motion, and when it is fixed, the model looks like a free-fermion model.

\section{\label{sec exact formulas}Reduced dynamics of a spin at $T=\infty$}
\subsection{Correlation function: sum over modes}

\begin{figure}[t]
\center{
\begin{tabular}{cc}
\includegraphics[width=0.5\textwidth]{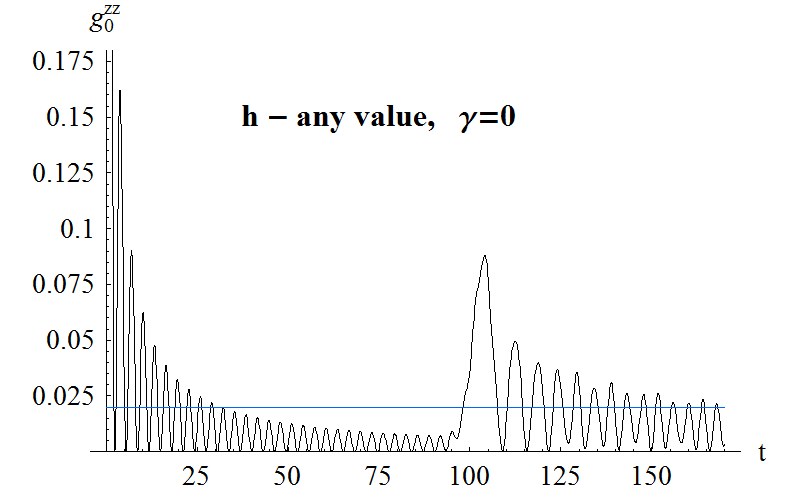} &     \includegraphics[width=0.5\textwidth]{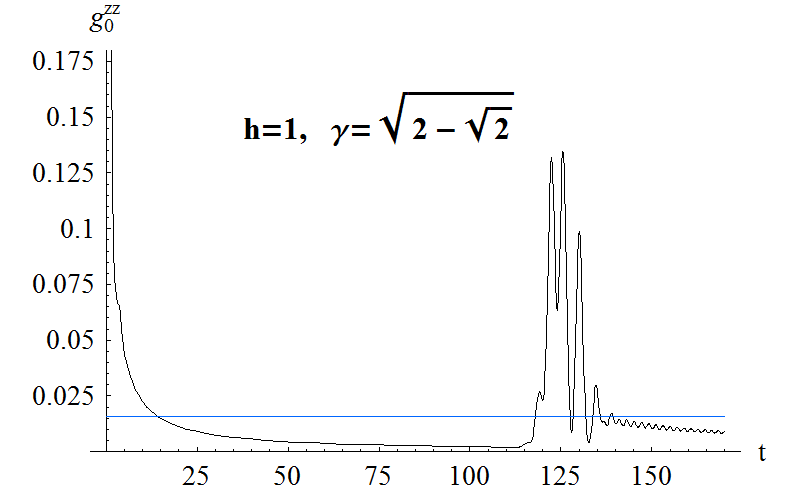} \\
\includegraphics[width=0.5\textwidth]{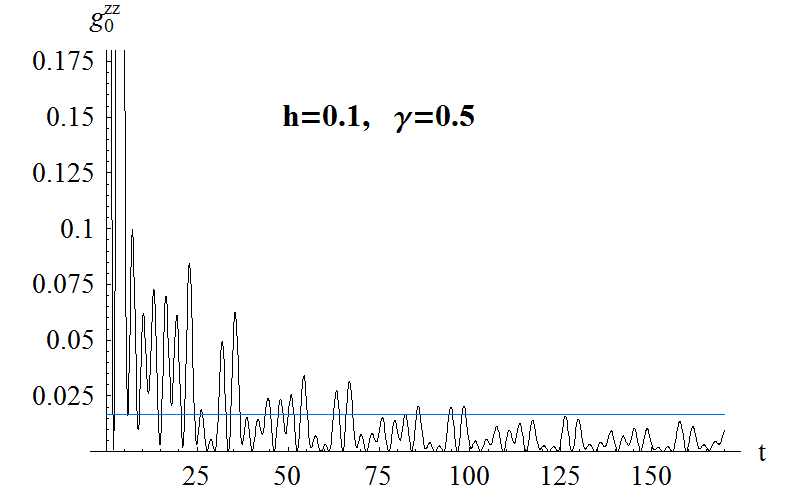} & \includegraphics[width=0.5\textwidth]{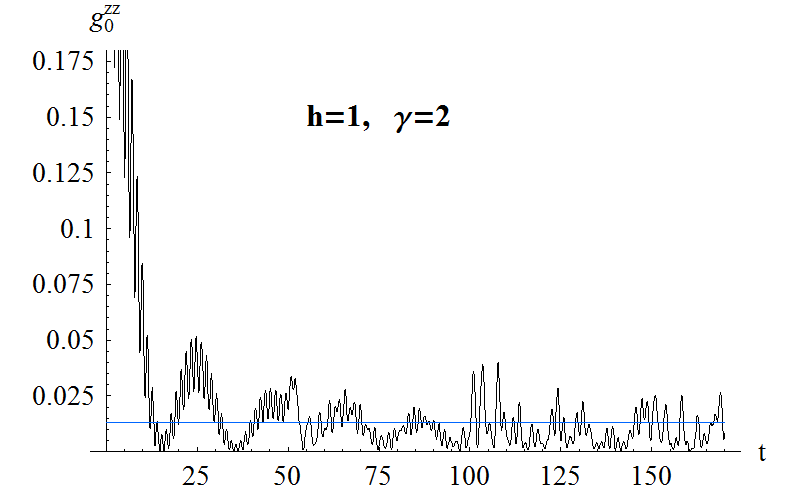} \\
\end{tabular}
}
\caption{(Color online)\label{fig variety of patterns} Patterns of spin dynamics for various values of model parameters. Exact autocorrelation function $g_0^{zz}(t)$ is plotted. According to eq. (\ref{polarization through correlation function}) it is  equal to the polarization of the first spin $p_1^z(t)$ provided $p_1^z(0)=1.$ 
Horizontal line (blue online)  marks the long-time average of the autocorrelation function given by eq.(\ref{long-time average}).
Number of spins here and in all other plots in the paper is $N=100.$
}
\end{figure}

We focus our study on the $z$-component of the $n$'th spin polarization vector as a function of time:
\be\label{polarization definition}
p^z_n(t)\equiv \tr[\rho(t) \sigma^z_n],
\ee
where $\rho(t)= e^{-i H t}\rho(0)e^{i H t}$ is the density matrix of the whole chain.

We choose the following initial condition:
\be\label{initial condition}
\dens(0)= 2^{-N} (\mathds{1}_1+\pp_1(0)\ssigma_1)\otimes  \mathds{1}_{23...N}.
\ee
It describes a situation when at $t=0$ the first spin has an arbitrary polarization $\pp_1(0),$ while other $(N-1)$ spins are completely unpolarized and uncorrelated. If the first spin is regarded as an open system, while other $(N-1)$ spins -- as an environment, then such initial condition corresponds to infinite temperature of the environment. Given the above initial condition, the polarization $p^z_n(t)$ can be expressed through the two-spin correlation functions at infinite temperature,
$
p_n^z(t)= p^\alpha_1(0) g^{z\alpha}_{n-1}(t),
$
where
\be \label{correlation function definition}
g^{z\alpha}_{n}(t)\equiv 2^{-N} \tr[ \sigma_{n+1}^\alpha(t) \sigma_1^z   ].
\ee
Due to conservation of parity $g^{zx}_n(t)=g^{zy}_n(t)=0$ and we are left with
\be\label{polarization through correlation function}
p_n^z(t)= p^z_1(0) g^{zz}_{n-1}(t).
\ee
Note that such relation between the polarization of a single spin and the correlation function holds only in the case of infinite temperature.

Thus our problem reduces to investigation of the $zz$ correlation function. Due to integrability of the model it may be calculated exactly \cite{mazur1973time,siskens1974time}. For the completeness of the presentation we provide the details of calculation in Appendix \ref{appendix Exact correlation function}. The result reads:



\be\label{pz(t) at beta=0}
g^{zz}_n(t) = \frac12 ({A^n_{\rm odd}}^2+{A^n_{\rm ev}}^2+{B^n_{\rm odd}}^2+{B^n_{\rm ev}}^2-{C^n_{\rm odd}}^2-{C^n_{\rm ev}}^2),
\ee
where
\be\label{pz(t) at beta=0, definitions}
\begin{array}{rcl}
A^n_{\rm ev (odd)}(t) & = &N^{-1}\sum\limits_{q\in X_{\rm ev (odd)}}\cos n\varphi(q) \cos E_q  t,  \\
B^n_{ \rm ev (odd)}(t) & = & N^{-1}\sum\limits_{q\in X_{\rm ev (odd)}}\frac{\varepsilon_q}{E_q}\cos n\varphi(q)  \sin E_q  t, \\
C^n_{ \rm  ev (odd)}(t) & = & N^{-1}\sum\limits_{q\in X_{\rm ev (odd)}} \frac{\Gamma_q}{E_q} \sin n\varphi(q) \sin E_q t.
\end{array}
\ee

In what follows we mainly concentrate on the evolution of the first spin which is distinguished by the initial condition. It is described by the autocorrelation function $g^{zz}_0(t).$

As was noticed in \cite{fel1999regular}, in case of the $XX$ model ($\gamma=0$)  $g^{zz}_n(t)$ is always non-negative (because \mbox{$C_{\rm  ev (odd)}^n(t)=0$}) or, in other words, spin polarization never changes its sign. We see that this is not the case for an arbitrary site $n$ in a general $XY$ chain. However, the polarization of the first spin still never changes its sign since \mbox{$C_{\rm  ev (odd)}^0 (t)=0$} for any $\gamma.$ Intriguingly, the same property (non-negativity of $g^{zz}_0(t)$ at infinite temperature) was observed in numerical simulations for the $XXZ$ model with long-range interactions \cite{bruschweiler1997non}. This suggests that this effect could be generic for a large class of spin systems.

Surprisingly enough, evolution of spin polarization described by the exact formula (\ref{pz(t) at beta=0}) exhibits a rich variety of patterns depending on $h$ and $\gamma.$ Examples are given in Fig. \ref{fig variety of patterns}. We aim at explaining major features of evolution and at providing tractable approximation to eq.(\ref{pz(t) at beta=0}).


\subsection{Correlation function: sum over winding numbers}
Let us rewrite formulae (\ref{pz(t) at beta=0, definitions}) for $n=0$
in a different form:
\be\label{pz(t) infinite sum}
\begin{array}{lcllcl}
 A_{\rm odd}^0(t) & = & A_0(t)+ 2\sum\limits_{j=1}^{\infty} A_j(t), & A_{\rm ev}^0(t) & = & A_0(t)+ 2\sum\limits_{j=1}^{\infty} (-1)^j A_j(t),\\
 B_{\rm odd}^0(t) & = & B_0(t)+ 2\sum\limits_{j=1}^{\infty} B_j(t), & B_{\rm ev}^0(t) & = & B_0(t)+ 2\sum\limits_{j=1}^{\infty} (-1)^j B_j(t),
\end{array}
\ee
where
\be\label{pz(t) at beta=0, definitions 2}
A_j(t)~~\equiv ~~(2\pi)^{-1} \Re \int\limits_{-\pi}^{\pi}e^{\ii(E(\varphi)t- j N\varphi)}d\varphi,
\ee
\be \nonumber
B_j(t)~~\equiv ~~(2\pi)^{-1} \Im \int\limits_{-\pi}^{\pi}\frac{\varepsilon(\varphi)}{E(\varphi)}e^{\ii(E(\varphi)t- j N\varphi)}d\varphi.
\ee
As will be shown below $j$ corresponds to a number of windings of a forefront of a wave packet produced by the initialization of the first spin.
To obtain the above expressions one should take a discrete Fourier transform of the r.h.s. of eq.(\ref{pz(t) at beta=0, definitions}) and use
\be
\begin{array}{rcl}
\sum\limits_{q\in X_{\rm odd}}e^{\ii l\varphi(q)} & = &
\left\{
\begin{array}{ll}
1 &{\rm if}~l=jN,~j\in Z,\\
0 & {\rm otherwise},\\
\end{array}
\right.
\\
\sum\limits_{q\in X_{\rm ev}}e^{\ii l\varphi(q)} & = &
\left\{
\begin{array}{ll}
(-1)^{j} &{\rm if}~l=jN,~j\in Z,\\
0 & {\rm otherwise}.\\
\end{array}
\right.
\end{array}
\ee

\begin{figure}[t]
\center{
\begin{tabular}{rl}
\includegraphics[width=0.5\textwidth]{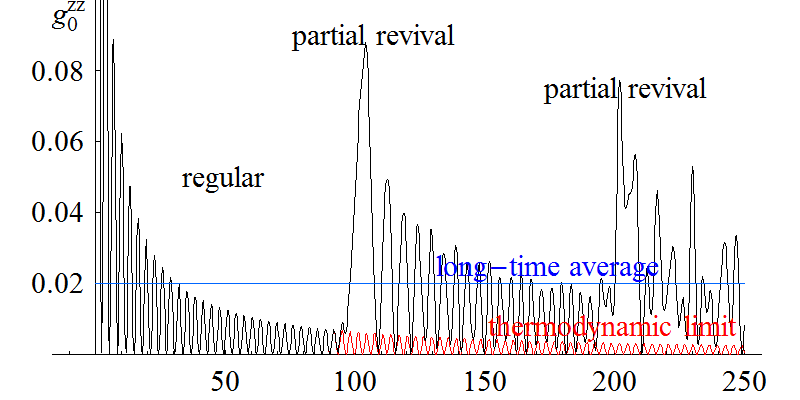}
\includegraphics[width=0.5\textwidth]{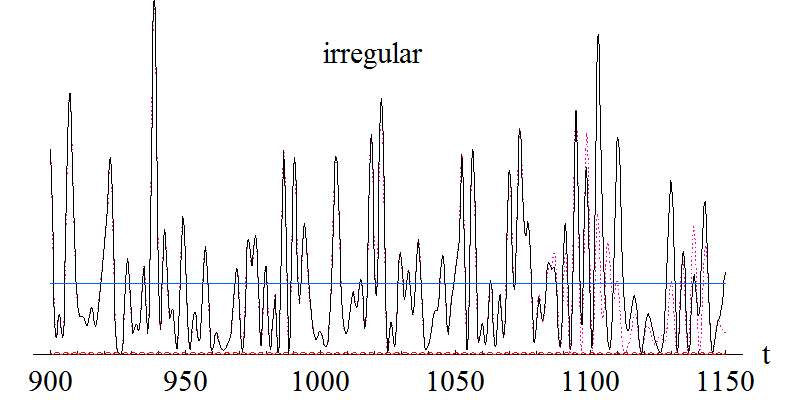}
\end{tabular}
}
\caption{(Color online)\label{fig XX} $g^{zz}_0(t)$ for the $XX$ model. Threshold time is $t_{\rm th}=N=100.$ Solid line corresponds to the exact expression. Dotted line (magenta online) corresponds to the approximation (\ref{pz(t) at beta=0 in XX successive approximations}) with $j=0,1,...,10.$  One can see that the approximation starts to deviate from the exact expression only at $t\simeq 11 t_{\rm th}.$ Approximation obtained in thermodynamic limit ($j=0$) is also shown (red online). It accurately describes $g^{zz}_0(t)$ up to the threshold time.  Horizontal line (blue online)  marks the long-time average of the autocorrelation function. }
\end{figure}

Formulae (\ref{pz(t) infinite sum}) have an important advantage compared to formulae (\ref{pz(t) at beta=0, definitions}): infinite sums in (\ref{pz(t) infinite sum}) may be truncated at some small $j$ to obtain excellent approximations for times $t<(j+1) t_{\rm th}$ with threshold time $t_{\rm th}\sim N.$ Thus one may deal with only few terms in eq. (\ref{pz(t) infinite sum}) in contrast to  $N$ terms in (\ref{pz(t) at beta=0, definitions}). This statement will be proved in full generality in what follows (see Sec. \ref{sec Asymptotics} and especially Appendix \ref{appendix t<jtTh}).  In two special cases described below one can check it immediately.
\subsection{Special case: $XX$ chain}


When $\gamma=0$ functions $A_j$ and $B_j$ can be expressed through Bessel functions of the first kind:
\be\label{A and B through Bessel}
\begin{array}{lcl}
A_j(t) & = & (-1)^{Nj/2} \cos(h t) J_{jN}( t), \\
B_j(t) & = & (-1)^{Nj/2} \sin(h t) J_{jN}( t).
\end{array}
\ee
$J_{jN}( t)$ is negligible for $t < jN,$ which justifies the truncation of the sums in (\ref{pz(t) infinite sum}). Threshold time in this case equals $N.$

In fact in the case of $XX$ chain eq. (\ref{pz(t) at beta=0}) may be further simplified to obtain
$$g^{zz}_0(t)~=  ~\frac{1}{2} \left(\left(J_{0}( t)+2\sum_{j=1}^\infty J_{jN}( t)\right)^2+\left(J_{0}( t)+2\sum_{j=1}^\infty (-1)^j J_{jN}( t)\right)^2\right)$$
\be\label{pz(t) at beta=0 in XX}
=\sum_{
j+j'=0({\rm mod}\, 2) 
}J_{jN}(t)J_{j'N}(t).
\ee
Note that $h$ drops out from the final expression. This can be easily seen from the definition~(\ref{correlation function definition}) of $g^{zz}_n(t)$ if one recalls that $\frac{h}2  \sum_{n} \sigma_n^z$ commutes with the total Hamiltonian.\footnote{As Prof. Perk noted in private communication, another way to explain this fact is to use the transformation into the rotating frame, a procedure familiar in the theory of magnetic resonance. }


Eq. (\ref{pz(t) at beta=0 in XX}) may be used to obtain successive approximations:
\be\label{pz(t) at beta=0 in XX successive approximations}
\begin{array}{l|l}
g^{zz}_0(t) \simeq & {\rm for}~t\in \\
\hline
J^{2}_{0}(t) & [0,N)\\
J^{2}_{0}(t)+4J^{2}_{N}(t)&  [N,2N)\\
J^{2}_{0}(t)+4J^{2}_{N}(t)+4J^{2}_{2N}(t)+4J_{0}(t)J_{2N}(t) & [2N,3N)\\
\multicolumn{2}{c}{...}
\end{array}
\ee
The first line ($j=0$) represents a well-known result obtained in thermodynamic ($N\rightarrow\infty$) limit \cite{niemeijer1967some}.
Approximations in which $(s+1)$ Bessel functions are kept correspond to  $s$ round trips of a spin wave over the circle.
We postpone further discussion of the physical sense of the obtained results to the next section.
Exact and approximate expressions for $g^{zz}_0(t)$ in the $XX$ chain are plotted at Fig. \ref{fig XX}.

Closely related results for the $XX$ model were obtained in ref. \cite{Zhu2010XX} and in refs. \cite{benderskii2011propagating,benderskii2013propagation}. In ref. \cite{Zhu2010XX} one-particle Green function (which is in fact equal to the {\it zero temperature} correlation function $g^{-+}_n(t)\left|_{T=0}\right.\equiv  \langle\downarrow\downarrow...\downarrow| \sigma_{n+1}^-(t) \sigma_1^+|\downarrow\downarrow...\downarrow\rangle$) was represented as an infinite sum  of Bessel functions. In refs. \cite{benderskii2011propagating,benderskii2013propagation} an open-ended $XX$ chain with an impurity in the center was considered, the impurity coupling being in general different from the bulk coupling; again the {\it zero temperature} autocorrelation function was represented as a sum over cycle number.

\subsection{Special case: Ising chain with $h=1$}
In the case $h=1,~\gamma=1$ one obtains
\be\label{A and B through Bessel h=1 gamma=1}
\begin{array}{lcl}
A_j(t) & = &  J_{2jN}( 2t), \\
B_j(t) & = &  \frac12 (J_{2jN+1}(2t)-J_{2jN-1}(2t))=-J_{2jN}'(2t),
\end{array}
\ee
where prime stands for the derivative.
Again $A_j(t)$ and $B_j(t)$ are negligible for $t < j t_{\rm th}$ with $t_{\rm th}=N.$ Successive approximations can be written analogously to the $XX$ case discussed above.

\section{\label{sec Asymptotics}Asymptotic approximations}

In the present section we derive asymptotic approximations for functions $A_j,B_j$ which enter eq.(\ref{pz(t) infinite sum}). As we will see, these approximations physically correspond to taking into account spin waves which wind over the circle $j$ times. The details of the calculations are presented in Appendix \ref{appendix Asymptotics}. Here we outline  only major results emphasizing their physical meaning. In the present section we restrict our study to the case $h\geq 1,$ $\gamma \in [0,1].$

\subsection{Winding of a wave packet over a circle}

\begin{table}[t]
\begin{center}
$
\begin{array}{|c||c|c|c|c|c|c|c|}
\hline
{\rm values ~of~}                     &   \gamma=0    &  \multicolumn{2}{|c|}{ \gamma=1 }   &   h=0                 & \multicolumn{2}{|c|}{ h=1 }  & h=1,          \\
\cline{3-4}
\cline{6-7}
{\rm parameters}
                   &               & h<1   & h\geq1                      &                       & \gamma^2\in[0,3/4) & \gamma^2\in[3/4,1] &  \gamma= \sqrt{2-\sqrt{2}} \\
\hline
\cos \varphi_0                      &     0         & h     &    1/h                      &    -\sqrt{\frac{\gamma}{1+\gamma}} & \frac{2\gamma^2+1-\sqrt{4\gamma^2+1}}{2(1-\gamma^2)} & 1       &    \sim 0.414     \\
\hline
V                                   & 1             & h     & 1                           & 1-\gamma   & -^* & \gamma  &      2(\sqrt{2}-1)                    \\
\hline
\end{array}
$
\end{center}
* {\small -- bulky (although explicit) expression}.
\caption{\label{table V} Value of maximal group velocity $V$ in some special cases}
\end{table}

We approximately calculate  $A_j(t)$ and $B_j(t)$ using the method of the steepest descent in the plane of complex variable $\varphi.$
The saddle points for $A_j$ are obtained from the equation
\be\label{saddle points}
v(\varphi)t-jN=0,
\ee
where $v(\varphi)\equiv \partial_\varphi E$ is the group velocity corresponding to momentum $\varphi$ (the equation corresponding to $B_j$ is slightly different, see eq.(\ref{general eq for saddle points}) in the Appendix \ref{appendix Asymptotics}). Note that in general the positions of saddle points depend on time  (to be more exact, on the ratio $t/jN$). Two important cases should be distinguished, $t< jt_{\rm th}$ and $t> jt_{\rm th},$ where $t_{\rm th}\equiv N/V$ and $V\equiv\sup\limits_{\varphi} v(\varphi)=v(\varphi_0).$ In the former case eq.(\ref{saddle points}) has no real roots and as a consequence $A_j(t)$ and $B_j(t)$ are severely suppressed (in accordance with a general result \cite{lieb1972finite}). This explains why one can keep only $j$ terms in eq.(\ref{pz(t) infinite sum}) whenever $t< jt_{\rm th}.$ If $h$ is not too close to $1,$ the suppression law reads
\be
A_j(t),B_j(t) \sim \exp[-{\rm const} \cdot jN  \left(\frac{jt_{\rm th}-t}{jt_{\rm th}}\right)^{\frac{3}{2}}],~~~t< jt_{\rm th},
\ee
where the constant is of order of one and depends on $h$ and $\gamma,$
see Appendix \ref{appendix t<jtTh}.

In the opposite case  $t> jt_{\rm th}$ eq.(\ref{saddle points}) has two real roots and $A_j(t),B_j(t)$ are not suppressed.

\begin{figure}[p]
\center{
\begin{tabular}{cc}
{spin polarization} & {spin polarization}\\
\includegraphics[width=0.5\textwidth]{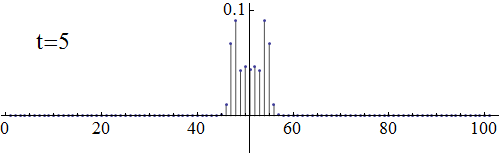} &
\includegraphics[width=0.5\textwidth]{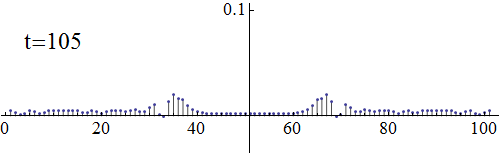} \\
\includegraphics[width=0.5\textwidth]{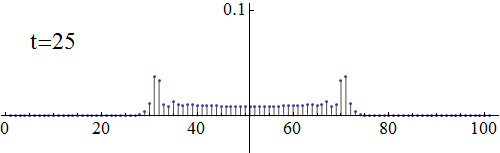} &
\includegraphics[width=0.5\textwidth]{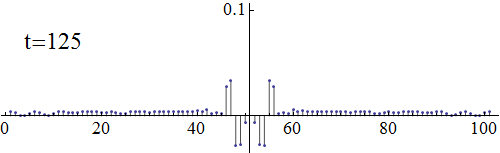} \\
\includegraphics[width=0.5\textwidth]{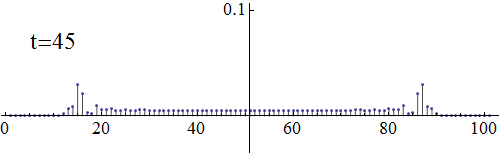} &
\includegraphics[width=0.5\textwidth]{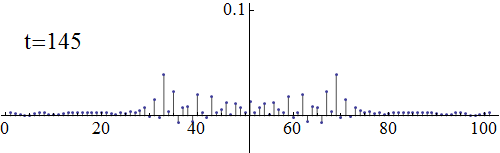} \\
\includegraphics[width=0.5\textwidth]{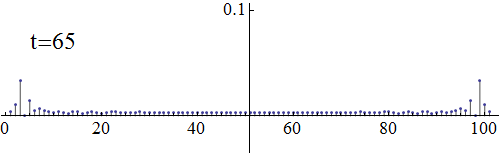} &
\includegraphics[width=0.5\textwidth]{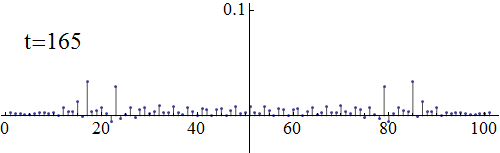} \\
\includegraphics[width=0.5\textwidth]{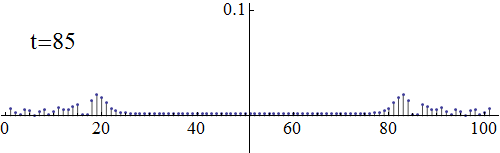} &
\includegraphics[width=0.5\textwidth]{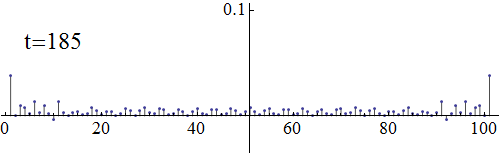}\\
spin site number & spin site number
\end{tabular}
}
\caption{(Color online)\label{fig whole chain} Propagation of two oppositely directed wave packets along the spin chain. Snapshots of polarizations of all spins at times $t=5,25,...185$ are presented. In this figure the initial spin excitation is localized at the $51$'st site (in contrast to the rest of the paper), the zeroth site is identified with the hundredth one.  The model parameters are $h=1,$ $\gamma=\sqrt{\sqrt2-1},$ the threshold time is $t_{\rm th} \simeq 117.7.$  Note the emergence of flat polarization plateau which grows up to $t_{\rm th}/2,$ then shrinks and completely disappears at threshold time. This feature is specific for the case $h=1.$ A three-dimensional plot representing  a more generic case (without plateau) of propagation of wave packets along the cyclic $XX$ chain can be found in \cite{fel1999regular}. }
\end{figure}

Threshold time $t_{\rm th}$ is a time which is necessary for the fastest spin wave to make one round trip over the circle \cite{fel1999regular}. Thus $A_j(t)$ and $B_j(t)$ describe contributions of those parts of the wave packet which have completed exactly $j$  round trips over the circle.  The propagation of the wave packet is visualized in Fig. \ref{fig whole chain} (see also an analogous figure for the $XX$ model in \cite{fel1999regular}).  As  was shown in \cite{fel1999regular}, initial excitation of the first spin gives rise to two wave packets which travel in opposite directions. Each wave packet is a superposition of all spin waves of corresponding direction.  The velocity of the forefronts of these wave packets coincides with the maximal group velocity of the spin waves $V.$  Therefore as long as $t<t_{\rm th}\equiv N/V,$ the wave packets propagate as if the chain were infinite, and the evolution of the first spin is described merely by oscillations in the common tail of the wave packets. This stage of evolution is the only one which may be catched by the $N\rightarrow\infty$ approximation. Mathematically it is described by keeping only $j=0$ terms in eq.(\ref{pz(t) infinite sum}).

At $t=t_{\rm th}$ the forefronts of two wave packets complete the round trip over the circle and meet at the first site. At this moment the regular evolution of the polarization of the first spin is abruptly interrupted by a partial revival. The succeeding evolution between $t_{\rm th}$ and $2t_{\rm th}$ is determined by the interference between the fastest parts of wave packets which have already made one round trip and the common tail of the wave packets with zero velocity which still stays at the first site. Mathematically this stage is described by keeping $j=0,1$ terms in eq.(\ref{pz(t) infinite sum}).

Subsequent stages are described in a similar fashion. The wave packets continue to wind over the circle. At  $s t_{\rm th} < t < (s+1) t_{\rm th}$ polarization at the first site is a result of interference of waves which completed $0,1,...,s$ round trips over the circle. This corresponds to keeping $j=0,1,..,s$ terms in eq.(\ref{pz(t) infinite sum}). The revivals at $t=s t_{\rm th}$ become less pronounced with increasing $s$ due to the decrease of the maximal amplitude and the smearing of the forefront  of the wave packet, see Fig. \ref{fig whole chain}.

Clearly the maximal group velocity $V$ is an important quantity in the above picture as it determines the threshold time. We show in Appendix \ref{appendix velocity} that $V \in [2(\sqrt{2}-1),1]$  as long as one restricts himself to the case $h \geq 1.$ Without this restriction $V$ is confined to the interval $[0,1].$ $V =2(\sqrt{2}-1)$ is achieved at $h=1,$ $\gamma=\sqrt{\sqrt{2}-1}$ (this case is presented on the upper right plot in Fig. \ref{fig variety of patterns}).  More detailed considerations including several important specific cases may be found in Appendix \ref{appendix velocity}.

\bigskip
The above described physical picture of propagation of wave packets and emergence of revivals implies that the forefront of the wave packet is rather sharp. This is indeed true for the following simple reason.
As long as $\varphi_0$ is a point of maximum, a bunch of fermions exists with $\varphi(q)$ lying in the vicinity of $\varphi_0.$ The group velocities of these modes are equal to each other and to the maximal velocity $V$ up to quadratic terms. It it is exactly these modes which form the sharp forefront of the wave packet which smears very slowly compared to the rest of the wave packet. Some proposals for high-quality quantum state transfer along spin chains exploit this feature (see e.g. \cite{Banchi2010}).

\subsection{\label{sec asymptotics 0}Asymptotic approximations for $t< t_{\rm th}.$}

\begin{figure}[t]
\center{
\includegraphics[width=\textwidth]{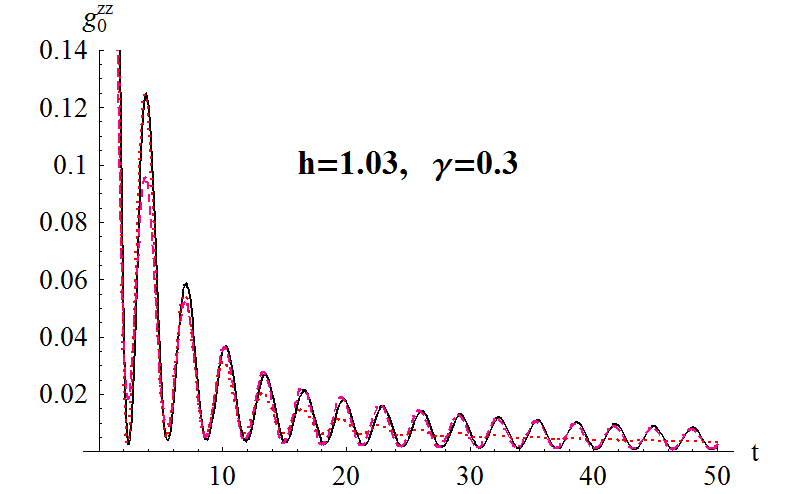}
}
\caption{(Color online)\label{fig approximation j=0} Autocorrelation function $g^{zz}_n(t)$ at $t<t_{\rm th}$ or, equivalently, in thermodynamic limit.  Solid line -- exact expression, dashed line (magenta online) -- asymptotic approximation (\ref{pz(t) in infinite XY 1}) valid for  lager times, $\epsilon^{-1} <t<t_{\rm th}$, dotted line (red online) -- asymptotic approximation (\ref{pz(t) in infinite XY 2}) valid for smaller times,  $1 < t \ll \gamma^{2}\epsilon^{-2}$.}
\end{figure}

First we separately consider the case $t<t_{\rm th}$ (see Appendix \ref{appendix t<tTh}). This case is special because the position of saddle points do not depend on time. Approximations at this time interval coincide with formulas obtained in thermodynamic limit.

Let us introduce a dimensionless parameter $\epsilon\equiv h-1.$ Here and in what follows we mainly concentrate on the case where $h$ is not too close to~$1.$ In this case the method of the steepest descend can be applied straightforwardly, and we obtain an asymptotic approximation for times $\max\{1,\epsilon^{-1}\} < t< t_{\rm th}:$

\be\label{pz(t) in infinite XY 1}
g^{zz}_0(t)\simeq \frac{1}{2\pi t}\cdot((a_{0+}-a_{0-})^{2}+4a_{0+}a_{0-}\cos^{2}(t -\frac{\pi}{4})),
\ee
where
\be \nonumber
 a_{0\pm}\equiv \sqrt{\frac{h \pm 1}{h \pm (1-\gamma^{2})}}.
\ee
Note that $\epsilon$ should be greater than $N^{-1},$ otherwise the time interval at which the approximation is valid vanishes.
This restriction is relaxed if $\gamma\ll\epsilon.$ In the latter case even for  $\epsilon\ll 1$ formula (\ref{pz(t) in infinite XY 1}) is valid for $1 \ll t<t_{\rm th}.$ In fact in this case one may approximate $g^{zz}_0(t)$ simply by the autocorrelation function for $\gamma=0$ which is given by $J_0^2(t)$ for $t<t_{\rm th}$ according to  (\ref{pz(t) at beta=0 in XX successive approximations}).

In the case of small $\epsilon$ the application of the method of the steepest descend is more sophisticated. The complications are not unexpected because $h=1$ is the point of quantum phase transition (QPT) for the $XY$ model. However given certain relations between $\epsilon,$ $\gamma$ and $N^{-1}$ one may still obtain accurate approximations, see Appendix \ref{appendix t<tTh}.   For example in the case $\epsilon^2 \lesssim \gamma^2  \ll 1$  we are able to obtain an asymptotic expression valid for $1 < t \ll \gamma^{2}\epsilon^{-2}:$
$$
g^{zz}_0(t)\simeq \frac{1}{ \pi(2-\gamma^2)t}
\left(
1+\frac{2-\gamma^2}{2\sqrt{1-\gamma^2}}\exp[-\frac{2\gamma^{2}}{\sqrt{1-\gamma^{2}}} t]
\right.
$$
\be\label{pz(t) in infinite XY 2}
\left.
+2\sqrt{\frac{2-\gamma^2}{2\sqrt{(1-\gamma^2)}}}\exp[-\frac{\gamma^{2}}{\sqrt{1-\gamma^{2}}} t] \cos(2t-\frac{\pi}{4}-\arctan\frac1{\sqrt{1-\gamma^2}})
\right)
\ee
If $\gamma$ is not too small, the exponents rapidly decrease with time and one is left with non-oscillating decay:
\be
g^{zz}_0(t) = \frac{1}{ \pi(2-\gamma^2)t}.
\ee

For certain values of parameters both asymptotic approximations (\ref{pz(t) in infinite XY 1}) and (\ref{pz(t) in infinite XY 2}) may be applicable, but at different time intervals.  An example is given in Fig. \ref{fig approximation j=0}.

\subsection{\label{sec asymptotics j}Asymptotic approximations for $t>jt_{\rm th}.$}

\begin{figure}[t]
\center{
\includegraphics[width=\textwidth]{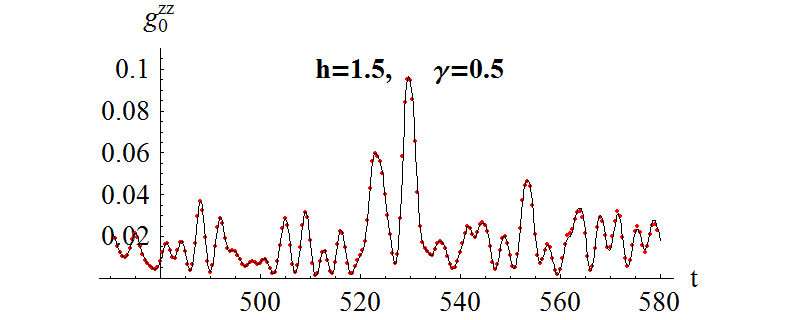}
}
\caption{(Color online)\label{fig approximation j>0} Exact autocorrelation function $g^{zz}_0(t)$ (solid line) and the approximation corresponding to 5 complete round trips over the circle (points). One can see that the approximation excellently describes both the revival and the irregular evolution far from the revival.
 }
\end{figure}

Now let us turn to asymptotics for functions $A_j(t)$ and $B_j(t)$ and corresponding approximations for $g^{zz}_0(t)$ in case of $j \geq 1.$ It was already noted that $A_j(t)$ and $B_j(t)$ are suppressed for $t<j t_{\rm th}.$

For sufficiently large times and  $h$ not too close to $1$ (more explicitly, for $(t-j t_{\rm th})\gg1$ and $\epsilon^{-1} \ll j t_{\rm th}$) we obtain (see Appendix \ref{appendix t>jtTh})
\be \label{formula for A asymptotic non-zero order}
A_j(t)\simeq \frac{1}{\sqrt{2\pi t}}\left(\sqrt{\frac{1}{|E''(\varphi_1)|}}\cos(tE(\varphi_1)-jN\varphi_1+\frac{\pi}{4})\right.
\ee
\be \nonumber
+\left.\sqrt{\frac{1}{|E''(\varphi_2)|}}\cos( tE(\varphi_2)-jN\varphi_2-\frac{\pi}{4})+O(\frac{1}{t})\right),
\ee
\be \label{formula for B asymptotic non-zero order}
B_j(t)\simeq \frac{1}{\sqrt{2\pi t}}\left(\sqrt{\frac{1}{|E''(\varphi_1)|}}\frac{ \varepsilon(\varphi_{1})}{E(\varphi_{1})}\sin(tE(\varphi_1)-jN\varphi_1+\frac{\pi}{4})\right.
\ee
\be \nonumber
\left.+\sqrt{\frac{1}{|E''(\varphi_2)|}}\frac{\varepsilon(\varphi_{2})}{E(\varphi_{2})}\sin( tE(\varphi_2)-jN\varphi_2-\frac{\pi}{4})+O(\frac{1}{t})\right).
\ee
Saddle points $\varphi_{1,2}(t)$ are obtained from Eq. (\ref{saddle points}). The latter may be reduced to a polynomial equation (\ref{eq on E'(z)}) of fourth degree with regard to $\cos\varphi.$

These asymptotics being plugged into eq.(\ref{pz(t) infinite sum}) excellently approximate  $g^{zz}_0(t)$ everywhere but in the vicinity of points $j t_{\rm th}$ where revivals occur. In order to describe revival one should use different approximate expressions presented in the next subsection.

The case of small $\epsilon$ is again more cumbersome. However it can also be treated as is demonstrated in Appendix \ref{appendix t>jtTh}.



\subsection{\label{sec revivals}Partial revivals
}

Above derived approximations based on the method of the steepest descent are not applicable in the vicinity of multiples of $t_{\rm th}$ when two saddle points are close to each other and to the point $\varphi_0.$ However, as long as at $t=j t_{\rm th}$ is exactly the time when a partial revival occurs, it is highly desirable to have an approximation which works well for $t\simeq j t_{\rm th}$. We present such an approximation in the Appendix \ref{appendix time near jt_th}. It is based on the fact that in the case under consideration the integrals in the definitions (\ref{pz(t) at beta=0, definitions 2}) of $A_j$ and $B_j$ pick up the major contribution in the vicinity of $\varphi_0.$ This justifies the expansion of $E(\varphi)$ in the vicinity of $\varphi_0$ which leads to the desired approximate expressions. If $h$ is not too close to $1,$ namely  $\epsilon\gg\gamma N^{-1},$  they read
\be \label{airy  Aj}
\begin{array}{cccl}
A_j(t) & \simeq & & \left(\frac{2}{|E'''(\varphi_0)|t}\right)^{\frac{1}{3}}\cdot \textrm{Ai}\left[-N\frac{t-jt_{\rm th}}{t_{\rm th}}\left(\frac{2}{|E'''(\varphi_0)|t}\right)^{\frac{1}{3}}\right]\cdot \cos(E(\varphi_0)t-jN\varphi_0)\\
B_j(t) & \simeq & \frac{\varepsilon(\varphi_0)}{E(\varphi_0)} & \left(\frac{2}{|E'''(\varphi_0)|t}\right)^{\frac{1}{3}}\cdot \textrm{Ai}\left[-N\frac{t-jt_{\rm th}}{t_{\rm th}}\left(\frac{2}{|E'''(\varphi_0)|t}\right)^{\frac{1}{3}}\right]\cdot \sin(E(\varphi_0)t-jN\varphi_0),
\end{array}
\ee
where $\textrm{Ai}(x)$ is the Airy function of the first kind and  $\varphi_0$ corresponds to the maximal group velocity. Curiously enough, the above approximation works well even far from $j t_{\rm th}.$  Note that as long as  $E'''(\varphi_0)$ does not depend on time in contrast to $E''(\varphi_{1,2}),$ it is much easier in practice to calculate the r.h.s. of equations (\ref{airy  Aj}) than the r.h.s. of eqs.  (\ref{formula for A asymptotic non-zero order}), (\ref{formula for B asymptotic non-zero order}). The only disadvantage of the approximation (\ref{airy  Aj}) is that we do not  analytically control the errors of this approximation; however, numerical calculations show that they are small.

To summarize, in order to approximate the autocorrelation function up to $(s+1)t_{\rm th},$ one should take $A_0,$ $B_0$ according to eq. (\ref{zero order spectral function asymptotic}), $A_j,$  $B_j$ with $j=1,2,...,s-1$ according to eqs. (\ref{formula for A asymptotic non-zero order}), (\ref{formula for B asymptotic non-zero order}) and $A_s,$  $B_s$ according to eq. (\ref{airy Aj}). The resulting expression  approximates the  autocorrelation function with excellent precision as shown in Fig. \ref{fig approximation j>0}.


The case when $h \simeq 1$ is as usual more cumbersome. We do not provide a complete analysis which would be rather bulky,  however we derive an approximation for $h=1, \gamma^2 \geq 3/4,$ see Appendix \ref{appendix time near jt_th}, eqs. (\ref{airy general integral for h=1 gamma2 near 075})--(\ref{airy h=1 gamma = 075 app Aj}).

\begin{figure}[t]
\center{
\includegraphics[width=\textwidth]{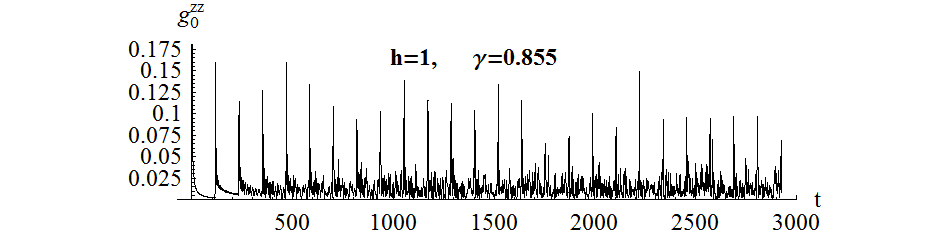}
}
\caption{\label{fig revivals} Extremely pronounced revivals occur in otherwise erratic regime in the small region of the parameter space  in the vicinity of the point $h=1,\gamma^2=3/4$. Threshold time here is $t_{\rm th} \simeq 117.$}
\end{figure}

Let us discuss the law which governs the decrease of revival amplitudes. In general the $s$'th partial revival is described  by the the mutual interference between all  $A_{j}(t)$ with $j=0,1,...,s$  and mutual interference between all  $B_{j}(t)$ with $j=0,1,...,s.$  However as a first approximation one may consider only $A_{s}(t)$ and $B_{s}(t)$ which give the leading contribution. $A_j(t)$ in eq.(\ref{airy Aj}) decreases as $t^{-1/3},$  while in eq. (\ref{formula for A asymptotic non-zero order}) -- as $t^{-1/2}.$  This means that the amplitude of revivals decreases more slowly than the averaged value of  $g^{zz}_{0}(t)$ between revivals. In fact this makes the revivals so visible against the background.  As a result for $h-1>\gamma/N$ one gets from eq.(\ref{airy  Aj}) the following law:
\be\label{revival decrease}
g^{zz}_{0}|_{s{\rm 'th~ revival}} \sim (s N)^{-{2}/{3}}
\ee
This law work satisfactory for sufficiently large number of spins and for moderate $s$. In particular, as long as the long-time average of $g^{zz}_{0}$ is of order of $N^{-1},$ this law can not be valid for $s\gtrsim \sqrt{N}.$ In fact it breaks down somewhat earlier because at large $s$ contributions from $A_j,~B_j$ with $j<s$ start to contribute significantly. Our numerical calculation show that for $10000$ spins the above law is reliable for a few dozens of revivals. For more moderate number of spins, $N\sim 100,$ the law is quickly distorted due to the above mentioned contribution from   $A_{j}(t)$ and $B_{j}(t)$ with $j<s.$ In particular,  maxima of revivals do not decrease monotonically in this case, see Fig. \ref{fig revivals}.

Noteworthy, when $h=1$ and $\gamma^2=3/4$ the amplitudes of the revivals decrease even more slowly than implied by eq. (\ref{revival decrease}), namely
\be
g^{zz}_{0}|_{s{\rm 'th~ revival}} \sim (s N)^{-{2}/{5}} , h=1,~\gamma^{2}={3}/{4},
\ee
see Appendix \ref{appendix time near jt_th} and especially  eq. (\ref{airy h=1 gamma = 075 app Aj}) for the details. Again this law works for sufficiently large number of spins. However the fact that  the $h=1,~\gamma^2=3/4$ point of a parameter space is a special one reveals itself already for modest $N\sim100$: the revivals appear to be especially pronounced in the vicinity of this point (see fig. \ref{fig revivals}), although they do not decrease monotonically due to the above discussed interference of $A_j,~B_j$ with different $j$. A very similar effect was observed previously in \cite{Banchi2011}. Namely, it was numerically discovered that when one initially polarizes a spin at the edge of an open-ended $XY$ chain and allows the excitation to propagate to another edge, the attenuation of the amplitude of a wave packet is minimal for $h=1,~\gamma\simeq 0.7.$

\section{\label{sec Chaos} Transition from regular to erratic evolution}

\begin{figure}[t]
\center{
\includegraphics[width=\textwidth]{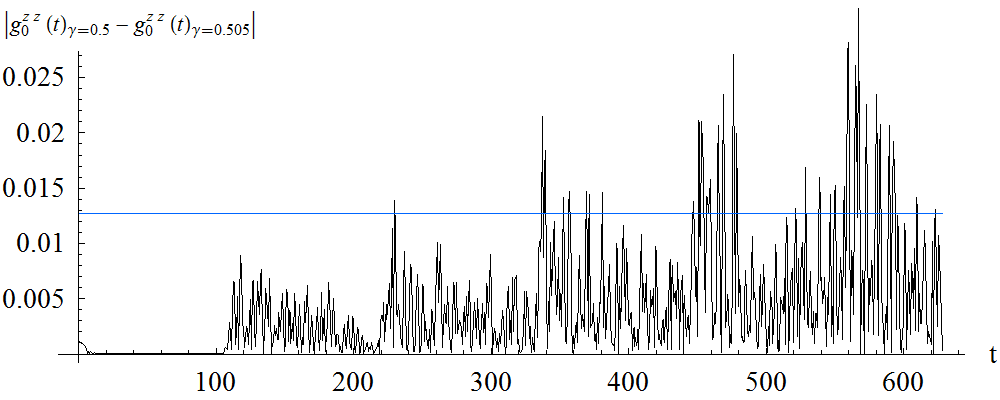}
}
\caption{(Color online)\label{fig chaos} Sensitivity of the correlation function to the small variation of $\gamma.$ Function plotted is the absolute value of difference of two correlation functions  $g^{zz}_0(t)$ corresponding to two slightly different values of anisotropy parameter: $\left| g^{zz}_0(t)_{\gamma=0.5}-g^{zz}_0(t)_{\gamma=0.505}\right|.$  Magnetic field in both cases is $h=1,$ number of spins $N=100.$ Threshold time is approximately $112$.
The horizontal line (blue online) shows the value which the above difference would admit if the two functions were absolutely uncorrelated.
}
\end{figure}

Plots of $g^{zz}_n(t)$ presented in the present paper clearly demonstrate that transition from regular to erratic evolution is a general feature of spin dynamics. In the present section we provide a discussion of this fact on a qualitative level. More thorough study including quantitative considerations will be presented elsewhere.

From Fig. \ref{fig XX}  it is evident that the spin evolution is apparently regular at small times but erratic (we are tempted to say "apparently chaotic") at large times, threshold time $t_{\rm th}$ determining the relevant timescale. However it is not so easy to define the terms "regular"$~$ and "erratic"$~$ rigorously  in the present context. One should especially be cautious when using the term $"$chaos$"$ here.
A widely used definition of quantum chaos is based on energy level repulsion (see e.g. \cite{haake2010quantum}). According to this
definition $XY$ model is certainly {\it not} chaotic because it is integrable and thus its level statistics is Poissonian (i.e. non-repulsive). In what follows we briefly discuss two distinct approaches which may be used to describe the level of irregularity of $g^{zz}_n(t).$

The first approach is based on the physical picture of winding of the wave packet over the circle and exploits asymptotic approximations derived above.
In this approach we pragmatically consider the evolution to be regular at some interval of time if the correlation function may be well approximated by a linear combination of {\it few} ($\ll N$) oscillating functions with different frequencies (probably multiplied by a power-law prefactor) at this interval. Conversely, the evolution is considered to be erratic when the approximation involves many ($\sim N$) harmonics. The first stage of evolution ($t<t_{\rm th}$) is the most regular one: according to eqs. (\ref{pz(t) in infinite XY 1}),(\ref{pz(t) in infinite XY 2}) it is described by a single cosine.
At times $t_{\rm th}, 2t_{\rm th}, 3t_{\rm th}$ new functions  $A_j,B_j$ come into play in eq.(\ref{pz(t) infinite sum}) and the number of harmonics increases stepwise. Thus the level of irregularity also increases. This does not last forever: according to eq.(\ref{pz(t) at beta=0}) $N$ harmonics is enough to describe $g^{zz}_{0}(t)$ {\it exactly.} Evidently the largest possible level of irregularity is achieved not later than at $t=Nt_{\rm th}.$ Note that here we use the term "harmonics"~ in a slightly non-standard way: we do not demand that the corresponding frequencies should be multiples of a single, minimal frequency. The described approach resembles the Feigenbaum rout to chaos through period doubling (see e.g. \cite{Feigenbaum}). However in the case under consideration there is no {\it doubling} -- the relation between frequencies of new harmonics switching on at certain times is not so evident. Moreover, these frequencies may even be slowly varying in time.  The Fourier analysis of the correlation function at different time intervals is necessary to obtain more quantitative picture. This will be done elsewhere.

In the second approach one examines the level of sensitivity of  $g^{zz}_n(t)$ to small variations of Hamiltonian parameters $h$ and $\gamma$. This approach introduced in \cite{peres1993quantum} resembles the definition of classical chaos through the extreme sensitivity to initial conditions.
We visualize the sensitivity of $g^{zz}_n(t)$ to small variations of $\gamma$ in Fig. \ref{fig chaos}. One can see that during the regular stage of evolution ($t<t_{\rm th}$)  such sensitivity is small, while at large times ($t>{\rm few} \cdot t_{\rm th}$) it is comparable to what one could expect if two correlation functions with slightly different parameters were absolutely mutually uncorrelated. As in the previous approach, the extent of thus defined irregularity increases stepwise at times which are multiples  of $t_{\rm th},$ the first step occurring at $t=t_{\rm th}$ being especially pronounced, see Fig. \ref{fig chaos}.
Curiously, our numerical experiments indicate that the sensitivity of $g^{zz}_n(t)$ to variations of $h$ and $\gamma$ generically tends to be  larger in the vicinity of quantum phase transition line $h=1.$  It would be quite surprising if this relation between QPT and sensitivity to small perturbations of Hamiltonian is confirmed, since the correlation function is calculated at infinite temperature while QPT occurs at zero temperature.

\section{\label{sec Conclusions} Summary}

 Numerical studies of the evolution of spin polarization in the finite cyclic XX chain \cite{fel1999regular} revealed the following physical picture (see also \cite{fel1998regular,Zhu2010XX,benderskii2011propagating,benderskii2013propagation}).

\begin{itemize}
\item  A threshold time exists up to which the polarization of a given spin evolves as if the chain were infinite. This is the time necessary for the fastest spin wave to make a round trip over the cyclic chain. Up to the threshold time the evolution is regular.
\item At the threshold time the regular evolution is interrupted by a partial revival. Subsequent partial revivals occur at $2t_{\rm th},3t_{\rm th},...$ Generically the evolution becomes more and more irregular (erratic) after each partial revival.

\end{itemize}

In the present paper we analytically justify this picture and generalize it to the anisotropic XY chain by developing a method to calculate the infinite-temperature correlation function for large times and beyond the thermodynamic limit.
Our core result is as follows.
\begin{itemize}
\item
We express the autocorrelation function $g^{zz}_{0}(t)$ as a series in winding number $j.$ An appealing feature of this representation is that the $j$th term does not contribute to the sum until $t=j t_{\rm th}$ and produces a partial revival  at $t=j t_{\rm th}.$ Each term of the series is defined in integral form. In two special cases ($\gamma=0$ and $\gamma=1,~h=1$) it can be expressed through Bessel functions. In a general case we provide very accurate explicit approximations valid at various times and in various regions of parameter space. Thus tractable approximations for $g^{zz}_{0}(t)$ at large times are obtained.
\end{itemize}
Other related results are as follows.
\begin{itemize}
\item
A parameter dependence of the threshold time $t_{\rm th}$ is analyzed.
\item
An asymptotic law of the revival amplitude decrease is established. This is a direct application of the above core result. For the bulk of the model parameter space the law has the form $\sim (j N)^{-{2}/{3}}$ (with $j$ being the number of the revival), however for special values of parameters it can be altered. In particular, in the vicinity of the point $h=1,~\gamma^2=3/4$ the law has the form $\sim (j N)^{-{2}/{5}}$ which leads to extremely pronounced revivals.\footnote{Such pronounced revivals were previously observed numerically in an open-ended chain \cite{Banchi2011}.}
\item
We show that a spin distinguished by the initialization retains the memory of this fact forever. In particular, its polarization (proportional to $g^{zz}_{0}(t)$) never changes the sign and  its time-averaged polarization differs from the time-averaged polarization of any other spin.\footnote{See  \cite{fel1999regular} for analogous conclusions in the context of the $XX$ model.} Thus we encounter the absence of the complete thermalization which is, however, only a finite-size effect (scaling as $1/N$).\footnote{ In thermodynamic limit the time-averaged polarization of any spin is zero, in agrement with both Gibbs distribution and {\it generalized} Gibbs distribution (which takes into account the integrals of motion, see e.g. \cite{LL-V} \cite{rigol2007relaxation}). The reason of this agreement is that our system is effectively at infinite temperature. Thus our work can not contribute to the ongoing debate on what is the correct equilibrium state of an integrable system}
\end{itemize}


A striking feature of the dynamics in the finite spin chain  is the transition from regular to erratic behavior. In the present paper we have restricted ourselves by brief and qualitative discussion of the nature and origin of this transition. Further work is necessary to give a more exhaustive and  quantitative  analysis.

\section*{Acknowledgements}
The authors acknowledge the enlightening comments by J.H.H. Perk and L. Banchi and the fruitful discussion at the Condensed Matter Theory seminar at ITAE RAS, especially valuable remarks made by A.L. Rakhmanov concerning signatures of onset of erratic behavior in the $XY$ model. O.L. also thanks  E. Bogomolny and O. Giraud for useful discussions. O.L. is grateful to ERC (grant no. 279738 – NEDFOQ) for financial support. The partial support from grants
NSh-4172.2010.2, RFBR-11-02-00778, RFBR-10-02-01398 and from the Ministry of Education and Science of the Russian Federation under contracts N$^{\underline{\rm o}}$N$^{\underline{\rm o}}$ 02.740.11.5158, 02.740.11.0239 is also acknowledged.

\appendix

\section{\label{appendix Diagonalization}Diagonalization of finite cyclic $XY$ spin chain}

\renewcommand{\theequation}{A-\arabic{equation}}
\setcounter{equation}{0}  


\subsection{Ranges of parameters}
Let us rewrite the Hamiltonian we are going to diagonalize:
\be\label{XY Hamiltonian in appendix}
H(h,\gamma,\kappa) =\frac\kappa4  \sum_{n=1}^N ((1+\gamma)\sigma_n^x \sigma_{n+1}^x+ (1-\gamma) \sigma_n^y \sigma_{n+1}^y)  + \frac{h}2  \sum_{n=1}^N \sigma_n^z,
\ee
where indices $1$ and $N+1$ are identified, and $N$ is even. Here we have introduced coupling constant $\kappa,$ which is taken to be $1$ everywhere in the article but this subsection.
Let us show  that one may consider  $h,\gamma,\kappa \geq 0$  without loss of generality. This means that one can change the sign of each constant by means of local unitary transformation $U.$ These transformations correspond merely to rotations of the coordinate systems at each spin site.

To change sign of $h$ one can transform  $\sigma_n^y\rightarrow-\sigma_n^y,$  $\sigma_n^z\rightarrow-\sigma_n^z$ at each spin site $n$:
$$
U=\prod_{n=1}^N  e^{i \sigma_n^x\pi/2}=\prod_{n=1}^N i \sigma_n^x,
$$
\be
U^\dagger \sigma_n^x U=\sigma_n^x,
~U^\dagger \sigma_n^y U=-\sigma_n^y,
~U^\dagger \sigma_n^z U=-\sigma_n^z,
~U^\dagger H(h,\gamma,\kappa) U=H(-h,\gamma,\kappa).
\ee
%
Analogously, to change sign of $\gamma$ one transforms  $\sigma_n^x\rightarrow\sigma_n^y,$  $\sigma_n^y\rightarrow-\sigma_n^x$ at each site $n$ by means of  $U=\prod\limits_{n=1}^{N} e^{i \sigma_n^z\pi/4}.$\\
To change sign of $\kappa$ one transforms  $\sigma_{2m}^x\rightarrow-\sigma_{2m}^x,$  $\sigma_{2m}^y\rightarrow-\sigma_{2m}^y$ at each {\it even} site $2m$ by means of  $U=\prod\limits_{m=1}^{N/2} e^{i \sigma_{2m}^z\pi/2}.$


As soon as sign of $\kappa$ is unimportant, one may put $\kappa=1.$

\subsection{$H$ in terms of $\sigma_n^\pm$}
We define the operators $\sigma_n^\pm$ in a usual way,
\be
\sigma_n^+=\frac12 (\sigma_n^x+i \sigma_n^y),~~~\sigma_n^-=\frac12 (\sigma_n^x-i \sigma_n^y).
\ee
These operators are neither Bose nor Fermi operators:
\be
\sigma_n^+ \sigma_n^- + \sigma_n^- \sigma_n^+=1,
\ee
\be
\sigma_m^+ \sigma_n^- = \sigma_n^- \sigma_m^+ {\rm~~~for~~~}m\neq n.
\ee
The following simple equalities prove to be useful:
\be
\sigma^z=2\sigma^+ \sigma^--1=-2\sigma^- \sigma^+ + 1
\ee
\be
\sigma^z\sigma^+=-\sigma^+\sigma^z=\sigma^+,~~~\sigma^z\sigma^-=-\sigma^-\sigma^z=-\sigma^-
\ee
The Hamiltonian may be rewritten in terms of $\sigma_n^\pm$ as follows:
\be
H=H_0+H_\gamma+H_h
\ee
with
\be
H_0 =\frac12 \sum_{n=1}^N (\sigma_n^+ \sigma_{n+1}^-+ \sigma_n^- \sigma_{n+1}^+),
\ee
\be
\hH_\gamma =\frac{\gamma}2 \sum_{n=1}^N (\sigma_n^+ \sigma_{n+1}^+ + \sigma_n^- \sigma_{n+1}^-),
\ee
\be
\hH_h =h \sum_{n=1}^N \sigma_n^+ \sigma_n^- -Nh/2.
\ee

\subsection{Jordan-Wigner transformation}
Define the operators
\be
\Pi_n \equiv \prod_{n=1}^n \sigma^z_n.
\ee
Evidently, $\Pi_N$ coincides with the parity operator $\Pi$ defined in Sec. \ref{sec XY model}.

Define Fermi operators $a^-_n,~a^+_n$ as follows
\be
a^-_n \equiv \sigma_n^- \Pi_{n-1} =\Pi_{n-1} \sigma_n^-,~~~ a^+_n \equiv \sigma_n^+ \Pi_{n-1} =\Pi_{n-1} \sigma_n^+
\ee
This implies
\be
\sigma_n^-= a^-_n  \Pi_{n-1}=   \Pi_{n-1} a^-_n,~~~\sigma_n^+= a^+_n  \Pi_{n-1}=  \Pi_{n-1} a^+_n ,
\ee

\be
\{a_m^+,a^-_n\}=\delta_{mn},~~~\{a_m^+,a^+_n\}=\{a_m^-,a^-_n\}=0,
\ee

\be
\sigma^z_n=2a^+_n a^-_n -1=-2a_n^- a_n^+ + 1.
\ee
The Hamiltonian takes the form (note that now the ordering of $a_n^\pm,~a_{n+1}^\pm$ is important; also note the change of the total sign):

\be
H_0 =- \frac12[ \sum_{n=1}^N (a_n^+ a_{n+1}^- + a_{n+1}^+a_n^-) - (1+\Pi)(a_N^+ a_1^- + a_1^+a_N^-)].
\ee

\be
\hH_\gamma =- \frac{\gamma}2[ \sum_{n=1}^N (a_n^+ a_{n+1}^+ + a_{n+1}^-a_n^-) - (1+\Pi)(a_N^+ a_1^+ + a_1^-a_N^-)].
\ee

\be
\hH_h =h \sum_{n=1}^N a_n^+ a_n^- -Nh/2.
\ee

\subsection{Fourier transformation}

Define for arbitrary real $q$
\be
b_q^-\equiv \frac{e^{i \pi/4}}{\sqrt N}\sum_{n=1}^N e^{-2 \pi i q (n-1)/N}a_n^-,~~~
b_q^+\equiv \frac{e^{-i \pi/4}}{\sqrt N}\sum_{n=1}^N e^{2 \pi i q (n-1)/N}a_n^+,
\ee
Then
\be\label{c anticommutation}
\{b_k^+,b^+_q\}=\{b_k^-,b^-_q\}=0,~~
\{b_k^+,b^-_q\}=\frac1N\frac{1-e^{2\pi i (k-q)}}{1-e^{2\pi i (k-q)/N}}.
\ee
In particular, if one takes
\be
q=-\frac{N}2+1,~-\frac{N}2+2,...,\frac{N}2 ~~~~~~ (X_{\rm odd})
\ee
or
\be
q=-\frac{N}2+\frac12,~-\frac{N}2+\frac32,...,\frac{N}2-\frac12 ~~~~~~ (X_{\rm ev})
\ee
then the set of $b_q^-$ is the set of Fermi annihilation operators.

The Hamiltonian may be written in terms of $b_q^\pm$ as follows:
\be
H=H^{\rm odd}P^{\rm odd}+H^{\rm ev}P^{\rm ev}
\ee
with
\be
P^{\rm odd}\equiv(1-\Pi)/2,~~~P^{\rm ev}\equiv(1+\Pi)/2,
\ee
\be
H^{\rm ev}=\sum_{q=1/2}^{N/2-1/2}~ H_q, ~~~ H^{\rm odd}=\sum_{q=1}^{N/2-1}~ H_q + H_{0,N/2},
\ee
\be
H_q=\left(h- \cos \varphi(q) \right)~(b^+_q b^-_q +b^+_{-q} b^-_{-q})
   +\gamma \sin \varphi(q)~ (b^+_q b^+_{-q}+b^-_{-q} b^-_q)-h,~~~\varphi(q) \equiv 2\pi q/N
\ee
and
\be
H_{0,N/2}=(h- 1)b^+_0 b^-_0 + (h+1)b^+_{N/2} b^-_{N/2}-h.
\ee





\subsection{Bogolyubov transformation}

Define the following quantities
\be
 \Gamma_q\equiv \gamma \sin \varphi(q), ~~~ \varepsilon_q\equiv h-\cos\varphi(q),~~~E_q\equiv\sqrt{\varepsilon_q^2+\Gamma_q^2},
\ee
Each $H_q$ may be written as follows:
\be
H_q=
(
\begin{array}{cc}
b^+_q & b^-_{-q}
\end{array}
)
\left(
\begin{array}{cc}
\varepsilon_q & \Gamma_q \\
\Gamma_q     & -\varepsilon_q
\end{array}
\right)
\left(
\begin{array}{c}
b^-_q \\ b^+_{-q}
\end{array}
\right).
\ee
It is possible to diagonalize this matrix through Bogolyubov transformation
\be\label{Bogolyubov transform}
c_q^-=\cos\frac{\theta_q}2 ~b_q^- + \sin \frac{\theta_q}2 ~b_{-q}^+.
\ee
The diagonalization condition reads $\tan \theta_q =\Gamma_q/\varepsilon_q,$ and we choose
\be\label{theta}
 \theta_q\equiv   \arctan\frac{\Gamma_q}{\varepsilon_q}~~{\rm for~~all~~}q\neq0.
\ee
This transformation preserves the anticommutation relations.
$H_{0,N/2}$ requires special treatment, which leads to $\theta_{N/2}=0,$
\be
\theta_0=
\left\{
\begin{array}{ll}
0,& h\geq1,\\
\pi, & 0\leq h<1.
\end{array}
\right.
\ee
The inverse transformation reads
\be\label{Bogolyubov transform}
b_q^-=\cos\frac{\theta_q}2 ~c_q^- - \sin \frac{\theta_q}2 ~c_{-q}^+
\ee
The odd and even parts of the Hamiltonian take the form
\be
H^{\rm odd ~(ev)}= \sum_{q\in X_{\rm odd~(ev)}}~ E_q (c^+_q c^-_q-\frac12).
\ee
This completes the diagonalization.

\subsection{Eigenstates}
Let us first prove the existence of the Fock vacuum states with respect to the annihilation operators $c_q^-,$ i.e. the states $|{\rm vac}\ra_{\rm odd},~|{\rm vac\ra_{\rm ev}}$ which satisfy
\be
c_q^-|{\rm vac}\ra_{\rm odd~(ev)}=0~~~\forall~q\in X_{\rm odd~(ev)}.
\ee
Evidently it is sufficient to prove that
\be\label{product operator is not zero}
\prod_{q\in X_{\rm odd ~(ev)}} c_q^- \neq 0.
\ee
If this condition is fulfilled,  one can always choose some states $|\Psi_{\rm odd(ev)}\ra$ and normalization constants $\aleph_{\rm ev}$ such that
\be
|{\rm vac}\ra_{\rm odd}=\aleph_{\rm odd} c_{-N/2+1}^-c_{-N/2+2}^-...c_{N/2}^-|\Psi_{\rm odd}\ra,
\ee
\be
|{\rm vac}\ra_{\rm ev}=\aleph_{\rm ev} c_{-N/2+1/2}^-c_{-N/2+3/2}^-...c_{N/2-1/2}^-|\Psi_{\rm ev}\ra.
\ee
The equality
\be
\{c_{-N/2+1}^+,[c_{-N/2+2}^+,\{...,\{c_{N/2-1}^+,[c_{N/2}^+\prod_{q\in X_{\rm odd }}c_q^-]\}...]\}=1
\ee
and the analogous equality for $q\in X_{\rm odd ~(ev)}$ prove eq.(\ref{product operator is not zero}).
Note that $|{\rm vac}\ra_{\rm ev}$ is indeed an eigenstate of the Hamiltonian, while $|{\rm vac}\ra_{\rm odd}$ is not.

All the eigenstates of the Hamiltonian are obtained from the vacuum states by applying the creation operators $c_q^+$. To create the odd number of fermions one should use $ q\in X_{\rm odd} $ and $|{\rm vac}\ra_{\rm odd}$, while to create the even number of fermions one should use  $q\in X_{\rm ev}$ and $|{\rm vac}\ra_{\rm ev}.$

Evidently one can enumerate all the eigenstates of the Hamiltonian by the multiindexes
\be
Q_M \equiv \{q_1,q_2,...,q_M\},~~~0 \leq M \leq N,
\ee
with the ordering $q_1<q_2<...<q_M.$ Then an eigenstate with $M$ fermions reads
\be
|Q_M\ra\equiv c^+_{q_M}...c^+_{q_2}c^+_{q_1}|{\rm vac}\ra_{\rm odd (ev)}
\ee
with $q_1,q_2,...,q_M\in X_{\rm odd~(ev)}$  when  $M$ is odd (even). The corresponding eigenenergy reads
\be
E_{Q_M}\equiv \sum_{q\in Q_M} E_q - \frac12 \sum_{q\in X_{\rm odd~(ev)}} E_q.
\ee

For our purposes we need only the matrix elements between the states with the same parity, therefore we use the notation $|{\rm vac}\ra$ without subscripts in what follows.

\section{\label{appendix Exact correlation function}Calculation of $g_n^{zz}(t)$}

  \renewcommand{\theequation}{B-\arabic{equation}}
  \setcounter{equation}{0}  

To calculate the  correlation function at infinite temperature,
\be \label{correletion function formula in App}
g^{zz}_{n}(t)=2^{-N}\sum_{Q,\tilde Q}\langle Q|\sigma_1^z|\tilde Q\rangle\langle\tilde Q|\sigma_{n+1}^z|Q\rangle e^{-i(E_{Q}-E_{\tilde Q})t},
\ee
one needs to calculate the corresponding matrix elements. To do this one uses
\begin{eqnarray}
a^+_{n+1}a^-_{n+1}=
\frac1N \sum_{p,\tilde p} &
\cos\frac{\theta_{\tilde p}}2  \sin\frac{\theta_p}2 c^+_{\tilde p} c^+_p e^{-2\pi i (p+\tilde p)n/N}+
\sin\frac{\theta_{\tilde p}}2  \cos\frac{\theta_p}2 c^-_{\tilde p} c^-_p e^{2\pi i (p+\tilde p)n/N}+ \nonumber\\
&
\cos\frac{\theta_{\tilde p}}2  \cos\frac{\theta_p}2 c^+_{\tilde p} c^-_p e^{2\pi i (p-\tilde p)n/N}+
\sin\frac{\theta_{\tilde p}}2  \sin\frac{\theta_p}2 c^-_{\tilde p} c^+_p e^{-2\pi i (p-\tilde p)n/N}.
\end{eqnarray}
Here $p,\tilde p$ can run {\it either} through  $X_{\rm odd}$ {\it or} through $X_{\rm ev}$ -- the expression is valid in both cases.
Now it can be easily seen that only three types of matrix elements do not vanish:

\begin{enumerate}
\item Diagonal matrix elements.
\be
\langle  Q|\sigma_{n+1}^z|Q\rangle =\frac1N\sum_{p\in X_M}\eta(Q_M,p)\cos\theta_p.
\ee
Here $\eta(Q_M,p)=1$ if $p\in Q_M$ and $-1$ otherwise; $X_M=X_{\rm odd~(ev)}$ if $M$ is odd (even).

\item Matrix elements between two states with equal number of fermions, differing by one fermion momentum.
\be\nonumber
Q_M=K_{M-1}\cup\{p\},~~~\tilde Q_M=K_{M-1}\cup\{\tilde p\}, ~~~ p,\tilde p \notin K_{M-1},~~p \neq \tilde p:
\ee
\be
\langle  \tilde Q| \sigma_{n+1}^z |Q\rangle=
e^{2\pi i(p-\tilde p)n/N} \frac2N \cos\frac{\theta_p+\theta_{\tilde p}}2 \langle  \tilde Q| c^+_{\tilde p} c^-_{p} |Q\rangle,
\ee
where $\langle  \tilde Q| c^+_{\tilde p} c^-_{p} |Q\rangle=\pm1,$ depending on the signature of the corresponding permutation. Note that this sign is not important for calculation of $g_n^{zz}(t)$.


\item Matrix elements between two states one of which can be obtained from another by addition of two fermions.
\be\nonumber
Q_M=\tilde Q_{M-2}\cup\{p\}\cup\{\tilde p\},~~~p,\tilde p \notin \tilde Q_{M-2},~~p \neq \tilde p:
\ee
\be
\langle  \tilde Q| \sigma_{n+1}^z |Q\rangle = \langle  Q| \sigma_{n+1}^z | \tilde Q\rangle^* =
-\frac2N e^{2\pi i(p+\tilde p)n/N} \sin\frac{\theta_p-\theta_{\tilde p}}2 \langle  \tilde Q| c^-_{\tilde p} c^-_{p} |Q\rangle.
\ee
\end{enumerate}

Now let us sum in eq. (\ref{correletion function formula in App}) separately over each type of matrix element:

\begin{enumerate}
\item
Sum over $Q=\tilde Q$ gives
$$
2^{-N}\left(
\frac{1}{N^{2}} \sum_{p,\tilde p\in X_{\rm odd}}  \cos\theta_p \cos\theta_{\tilde p}
\sum_{{\rm odd}~M } \sum_{Q_M} \eta(Q,p)\eta(Q,\tilde p)
+\{{\rm odd}\rightarrow{\rm even}\}
\right)
$$
\be \label{ME 1 contribution}
=\frac{1}{2N^{2}}\left(\sum_{p\in X_{\rm odd}}+\sum_{p\in X_{\rm ev}}\right) \cos^{2}\theta_{p}.
\ee

\item
Sum over pairs $(Q,\tilde Q)$ of the form
$Q=K\cup\{p\},~\tilde Q=K\cup\{\tilde p\}$ gives
\be\label{ME 2 contribution}
\frac{1}{2N^{2}}
\left(
\sum_{\substack{ p,q\in X_{\rm odd}\\ p \neq \tilde p}}+
\sum_{\substack{ p,q\in X_{\rm ev}\\ p \neq \tilde p}}
\right)
e^{2\pi i (p-\tilde p) n /N-i(E_{p}-E_{\tilde p})t}
\cos^2\frac{\theta_{p}+\theta_{\tilde p}}{2}.
\ee

\item
Sum over pairs $(Q,\tilde Q)$ of the form
$Q=K\cup\{p\}\cup\{\tilde p\},~\tilde Q=K$ gives
\be \label{ME 3 contribution}
\frac{1}{4N^{2}}
\left(
\sum_{\substack{ p,q\in X_{\rm odd}}}+
\sum_{\substack{ p,q\in X_{\rm ev}}}
\right)
e^{2\pi i (p+\tilde p) n/N-i(E_{p}+E_{\tilde p})t}\sin^{2}\frac{\theta_{p}-\theta_{\tilde p}}{2},
\ee
while summation over $Q=K,~\tilde Q=K\cup\{p\}\cup\{\tilde p\}$ gives a  complex conjugated contribution.
\end{enumerate}

If one takes $p=q$ in expression (\ref{ME 2 contribution}), it becomes equal to the (\ref{ME 1 contribution}) contribution. Exploiting this one readily obtains

\begin{eqnarray}\label{correlation function not factorized}
g^{zz}_{n}(t)= &
\frac1{2N^2}
\left(
\sum\limits_{\substack{ p,q\in X_{\rm odd}}}+
\sum\limits_{\substack{ p,q\in X_{\rm ev}}}
\right) &
\left(
\cos\left(\frac{2\pi(p-\tilde p) n}N-(E_{p}-E_{\tilde p})t\right)\cos^2\frac{\theta_{p}+\theta_{\tilde p}}{2}+ \right. \nonumber \\
& & \left. \cos\left(\frac{2\pi(p+\tilde p) n}N-(E_{p}+E_{\tilde p})t\right)\sin^2\frac{\theta_{p}-\theta_{\tilde p}}{2}
\right)
\end{eqnarray}
It can be straightforwardly verified that this expression leads to eqs. (\ref{pz(t) at beta=0}),(\ref{pz(t) at beta=0, definitions}).

Eq. (\ref{correlation function not factorized}) can be used to find a long-time average of the autocorrelation function $\overline {g^{zz}_{0}}\equiv \lim\limits_{T \rightarrow \infty} T^{-1}\int\limits_0^Tg^{zz}_{0}(t)dt.$ Let us assume that there are no degeneracies in $E_q$ other than the mirror degeneracy $E_q=E_{-q}$ (in other words, that $E_q=E_p$ implies $|q|=|p|$). This is a generic case. Then
\be\label{long-time average}
\overline {g^{zz}_{0}}=\frac1N(1-\frac1N+\frac1{2N}\left(
\sum\limits_{\substack{ p\in X_{\rm odd}}}+
\sum\limits_{\substack{ p\in X_{\rm ev}}}
\right)\cos^2\theta_p
).
\ee
Term $-1/N$ in parenthesis emerges due to $q=0,N/2.$ Long-time average of correlation function for $n\neq0$ can be calculated analogously. In the specific case $\gamma=0$ the result (\ref{long-time average}) coincides with the expression obtained in \cite{fel1999regular}.

\section{\label{appendix velocity} Group velocity of spin waves}

  \renewcommand{\theequation}{C-\arabic{equation}}
  \setcounter{equation}{0}  

In the present section we consider $h\geq0,$ $\gamma\in[0,1].$
Group velocity of spin waves reads
\be
v(\varphi;h,\gamma)=(h-(1-\gamma^2)\cos\varphi)\sin\varphi/E(\varphi;h,\gamma).
\ee
We are interested mainly in the {\it maximal} velocity for given values of parameters $h$ and $\gamma:$
\be
V(h,\gamma)\equiv\sup\limits_\varphi v(\varphi;h,\gamma)=v(\varphi_0;h,\gamma),
\ee
where $\varphi_0$ is the  supremum point. Due to the symmetry of $E(\varphi)$ we can consider $\varphi\geq 0$ without loss of generality. Extremum condition $\partial_\varphi v|_{\varphi_0}=0$  leads to the fourth degree polynomial equation
\be\label{fourth degree polynomial equation}
P(z) \equiv (1-b)^{2}z^{4}-3h(1-b)z^{3}+(2b(1-b)+h^{2}(3-2b))z^{2}-h(h^2+b)z-b(1-b)+h^{2}b=0
\ee
with  $z= \cos \varphi_0 $ and $b\equiv\gamma^2.$
We are interested in the real roots of this equation which lie in the interval $[-1,1].$ Let us show that there is only one such root whenever $h\geq1$ (this fact is important for the application of the method of the steepest descent, see Appendix \ref{appendix Asymptotics}).  In this case the above equation implies that $z \geq 0,$ therefore in fact we have to consider the interval $[0,1].$ Since $P(0)>0$, $P(1)<0$ and $P(+\infty)=+\infty$, we could have 1,2 or 3 roots in $[0,1]$. If there were 2 or 3 roots of $P=0$ in the considered interval, then the equation $P'=0$ would have 2 roots in $[0,1]$. However the latter equation has no more than 1 root in the considered interval (z=$\frac{h}{4(1-b)}$). Thus equation (\ref{fourth degree polynomial equation}) has exactly 1 root in the interval $[0;1]$ for $h\geq1$.

Let us now consider several important special cases.\\
{\bf I. $\gamma=0.$}
In this case $\varphi_0=\pi/2,$ $V=1.$\\
{\bf II.} {$\gamma=1.$}
In this case
\be
\cos\varphi_0=\left\{
\begin{array}{ll}
h, & h\leq1\\
h^{-1}, & h>1,
\end{array}
\right.
\ee
and
\be
V=
\left\{
\begin{array}{ll}
h, & h\leq1\\
1, & h>1,
\end{array}
\right.
\ee
\\
{\bf III.} {$h=0.$}
In this case $\cos\varphi_0=-\sqrt{\frac{\gamma}{1+\gamma}},$ $V=1-\gamma.$\\
{\bf IV.} {$h=1.$}
This is an especially interesting case as it corresponds to the quantum phase transition. Velocity $v(\varphi)$ has a step at $\varphi=0,$ step height being equal to $2\gamma.$ Eq. (\ref{fourth degree polynomial equation}) is simplified to
\be
(z-1)^2\left( (b-1)^2 z^2 + (2b^2-b-1)z + b^2 \right)=0.
\ee
One should distinguish two cases.\\
{\bf IV a.} $b \in [3/4,1].$ In this case the only root that satisfy $|z|\leq 1$ is $z=1.$ Thus $\varphi_0=0,$ $V=\gamma.$\\
{\bf IV b.} $b \in [0,3/4).$ In this case $\cos\varphi_0=\frac{2b+1-\sqrt{4b+1}}{2(1-b)}$  from which one can easily write down an expression for $V$ which appears to be somewhat bulky.  One can also find a {\it minimal} value of $V$ with respect to $\gamma:$
\be\label{minimal V}
\inf\limits_{\gamma\in[0,1]} V(1,\gamma)= V(1,\sqrt{2-\sqrt{2}})=2(\sqrt{2}-1)=0.828427...
\ee
In what follows we show that this is the minimal value of $V$ in the whole region $h\geq1, \gamma\in[0,1].$\\
{\bf V.} $h\rightarrow\infty.$ In this case $\varphi_0\rightarrow 0,$ $V\rightarrow 1.$

Let us investigate how $V(h,\gamma)$ varies with $h.$ The derivative over $h$ has a rather simple form:
\be
\partial_h V(h,\gamma)= b (1-h\cos\varphi_0)\sin\varphi_0/E^3(\varphi_0;h,\gamma).
\ee
To calculate it we used that $\partial_h V(h,\gamma)=\partial_h v(\varphi_0;h,\gamma)$ due to the equation \mbox{$\partial_\varphi v (\varphi_0;h,\gamma)=0.$} The stationary points of $V(h,\gamma)$ with respect to $h$ are given by $\partial_h V=0$ which leads to $\cos\varphi_0=1/h.$ We plug the latter equality into eq. (\ref{fourth degree polynomial equation}) and obtain
\be
(1-\gamma^2)(1-z^2)^2(1-(1-\gamma^2)z^2)=0.
\ee
The only roots that satisfy $|z|\leq 1$ are $z=\pm 1$ which correspond to $h=1.$ This point is not extremal because $V(0,\gamma)\leq V(1,\gamma)\leq V(+\infty,\gamma).$ Thus for any fixed $\gamma$ maximal group velocity $V(h,\gamma)$ monotonically grows with $h$ from $1-\gamma$ at $h=0$ to $1$ as  $h\rightarrow\infty.$ As a consequence, if one considers only $h\geq1,$ than the minimal value of $V$ is given by eq. (\ref{minimal V}).

\section{\label{appendix Asymptotics}Asymptotic expressions}

  \renewcommand{\theequation}{D-\arabic{equation}}
  \setcounter{equation}{0}  

Here we consider in detail asymptotic expressions for spectral functions $A_j(t)$ and $B_j(t)$. Let us explore domains in which $E(\varphi)$ is an univalent analytical function. The branch points of this function are found from the equation
$(h-\cos\varphi)^{2}+( \gamma \sin\varphi)^{2} = 0.$
For $h^{2}>1-\gamma^{2}$ its solutions are $\varphi_{br\pm}$ and $\varphi^{*}_{br\pm}$, where
\be
\varphi_{br\pm}=2\pi k + i\cdot \textrm{arccosh}\left(\frac{ h \pm \gamma  \sqrt{h^{2}-(1-\gamma^{2})}}{1-\gamma^{2}}\right),~~~~~k \in Z.
\ee
The domain where $E(\varphi)$ remains univalent analytical function is the hole complex plane without row of branch cuts from $\varphi_{br-}$ to $\varphi_{br+}$ in the $\im(\varphi)>0$ half-plane and row of branch cuts from $\varphi^{*}_{br-}$ to $\varphi^{*}_{br+}$ in the  $\im(\varphi)<0$ half-plane. It is important that $\im(E(\varphi))$ is positive in I and III quadrants and negative in II and IV quadrants. Functions $E(\varphi)$, $\varepsilon(\varphi),$  $\exp[-ijN\varphi],$ as well as integrands in eq.(\ref{pz(t) at beta=0, definitions 2}), are $2 \pi $ periodic functions of $\re (\varphi).$ Thus we consider the strip $\re(\varphi) \in (-\pi;\pi].$

In order to use the method of the steepest descent for the integrals in eq.(\ref{pz(t) at beta=0, definitions 2}) we have to find saddle points for functions
\be
f_{Aj}(\varphi) \equiv iE(\varphi)t-ijN\varphi,~~~~~~f_{Bj}(\varphi) \equiv iE(\varphi)t-ijN\varphi +\ln \frac{\varepsilon(\varphi)}{E(\varphi)}.
\ee
The saddle points are defined by the equations
\be \label{general eq for saddle points}
E' t-jN=0 ,~~~~~~\textrm{and}~~~~~  iE' t-ijN+\left( \ln\frac{\varepsilon}{E}\right)'=0
\ee
correspondingly.

\subsection{\label{appendix t<tTh}Asymptotics for spectral functions of zero order}

Let us first consider spectral functions of zero order, $A_{0}(t)$ and $B_{0}(t)$. There are four saddle points in the strip $\re(\varphi) \in (-\pi;\pi]$ which for $A_0$ read
\be
\varphi_{1}=0,~~\varphi_{2}= \pi,~~\varphi_{3}=-i\cdot \textrm{arccosh}\frac{h}{1-\gamma^{2}},~~\varphi_{4}=+i\cdot \textrm{arccosh}\frac{h}{1-\gamma^{2}}.
\ee
For $B_{0}$ we have exactly the same points $\varphi^{B}_{1}=\varphi_{1},$  $\varphi^{B}_{2}=\varphi_{2}$ and slightly (for $t \gg 1$) shifted points $\varphi^{B}_{3}$ and $\varphi^{B}_{4}$:
\be \label{saddle points for B_0}
\varphi^{B}_{3,4}= \mp i\cdot \textrm{arccosh}\left(\frac{h}{1-\gamma^{2}}+c_{3,4}\cdot\frac{1}{t}+O(\frac{1}{t^{2}})\right)
\ee
One can find $c_{3,4}$ by substituting $\varphi^{B}_{3,4}=\varphi_{3,4}+\vartriangle \varphi$ in the second equation in (\ref{general eq for saddle points}). Note that if in the  $(c_{3,4} \cdot t^{-1})$-vicinity of $\varphi_{3,4}$ the series for $E(\varphi)$ is convergent, then the difference between $\varphi_{3}$ and $\varphi^{B}_{3,4}$ is of order of $t^{-1}.$

\begin{figure}
\center{
\begin{tabular}{lr}
\includegraphics[width=0.5\textwidth]{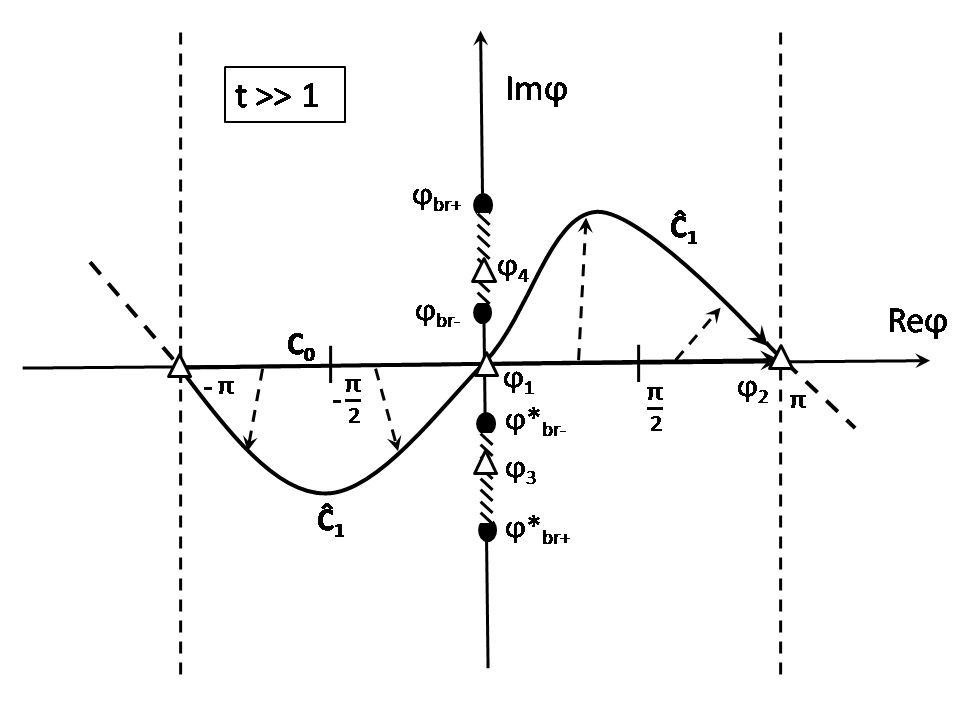} &
\includegraphics[width=0.5\textwidth]{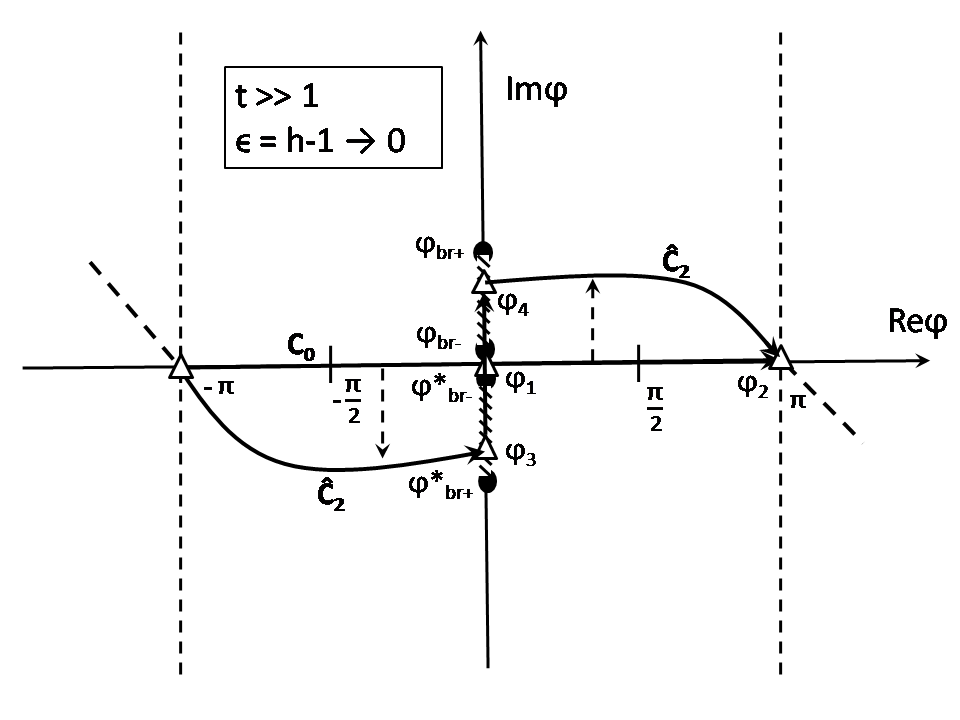}
\end{tabular}
}
\caption{\label{asyptotic zero order} Integration paths for $A_{0}$, $B_{0}$ in the case $t \gg 1,~ t \gg \epsilon^{-1}~$ (left) and in the case $\epsilon^{-1} \gg t \gg 1, ~ \gamma \gg \epsilon~$ (right). Empty triangles correspond to saddle points, filled circles -- to branch points of $E(\varphi)$.}
\end{figure}

Let us define the following parameter:
\be
\epsilon \equiv h-1.
\ee
For large enough $\epsilon$ we can find asymptotics in the case $t \gg \max \{\epsilon^{-1},1\}$ in a straightforward way. Indeed, we can transform the integration path from our initial $C_{0}$ (integration along $\re(\varphi)=0$ from $\varphi = - \pi$ to $\varphi = \pi$) to the path $\widetilde{C}_{1}$ which goes through I and III quadrant, where $\im(E)>0,$ and through saddle points $\varphi_{1}=0$ and $\varphi_{2}=\pi$ (see figure (\ref{asyptotic zero order}), left). Then we immediately have
\be \label{zero order spectral function asymptotic}
\begin{array}{rcl}
A_{0}(t) & \simeq & \sqrt{\frac{1}{2 \pi t}}\left(a_{0-} \cos((h-1)t+\frac{\pi}{4})+a_{0+}\cos((h+1)t-\frac{\pi}{4})+O(\frac{1}{t})\right),\\
B_{0}(t) & \simeq & \sqrt{\frac{1}{2 \pi t}}\left(a_{0-} \sin((h-1)t+\frac{\pi}{4})+a_{0+}\sin((h+1)t-\frac{\pi}{4})+O(\frac{1}{t})\right),
\end{array}
\ee
where
\be \nonumber
a_{0\pm}= \sqrt{\frac{h \pm 1}{h \pm (1-\gamma^{2})}}
\ee
and we take into account that $\varepsilon(0)E^{-1}(0)=\varepsilon(\pi)E^{-1}(\pi)=1$.
Region of applicability for large enough $\epsilon$ is given by $t \gg 1$, and for $\epsilon \rightarrow 0$ we have $t > 2\epsilon^{-1}\ln\frac{1}{\vartriangle}$ where $\vartriangle$ stays for the value of an error. Under these conditions we have the following approximation for $t<t_{\rm th}$:
\be\label{pz(t) in infinite XY}
g^{zz}_{0}(t)\simeq \frac{1}{2\pi t}\cdot ((a_{0+}-a_{0-})^{2}+4a_{0+}a_{0-}\cos^{2}(t -\frac{\pi}{4})).
\ee
Numerical evolution shows an excellent coincidence with the exact solution in the region $1 \ll t < t_{\rm th}$. Note that this expression becomes asymptotic for $J^{2}_{0}(t)$ in the case $\gamma=0~$ (XX chain) which is in accordance with eq. (\ref{pz(t) at beta=0 in XX successive approximations}).

Let us consider the case of such small $\epsilon$ that $\epsilon \ll \gamma$ and $\epsilon \ll (t)^{-1}$. Since we are interested in the dynamics on time scale of order of $t_{\rm th}$ or larger, the latter condition in fact implies
$\epsilon \ll N^{-1}\ll1$. Now we cannot integrate over the contour $\widetilde{C}_{1}$  due to the small convergence radius of the series in the vicinity of  $\varphi_1 = 0$ (note however that the contribution from the saddle point $\varphi_2=\pi$ remains intact).   Therefore we use another integration path, $\widetilde{C}_{2}$ (see fig. (\ref{asyptotic zero order})), which goes through saddle points $-\pi,$ $\varphi_{3}$, $\varphi_{1}$, $\varphi_{4}$ and ends up in $\varphi_{2}=\pi$. Consider integration path from $\varphi_{3}$ to $\varphi_{4}$. Integration over branch cuts does not contribute to $A_{j}$, $B_{j}$ (this statement is true for spectral functions of all orders). This is because $\re(E(\varphi))=0 $ on the branch cuts, and therefore
\be \label{integration through incision}
\re\int_{\rm cut}e^{itE(\varphi)-in\varphi}d\varphi=
\im\int_{\rm cut}\frac{\varepsilon(\varphi)}{E(\varphi)}e^{itE(\varphi)-in\varphi}d\varphi=0.
\ee
For small $\epsilon$ branch points can be expanded as $\varphi_{br-},\varphi_{br-}^{*}=\pm i \cdot \epsilon \gamma^{-1}+O(\epsilon^{2}\gamma^{-2})$. At the segment $[\varphi^{*}_{br-},\varphi_{br-}]$ functions $E(\varphi)$ and $\varepsilon(\varphi)$ are real-valued, and $E(\varphi)<\sqrt{2} \epsilon,~\varepsilon(\varphi)E^{-1}(\varphi)=1+O(\epsilon)$, therefore
\be \label{for zero order we neglect part of integral }
\re\int_{\varphi^{*}_{br-}}^{\varphi_{br-}}e^{itE(\varphi)}d\varphi~ <~(\sqrt{2} \gamma^{-1}\epsilon+O(\epsilon^{2}\gamma^{-2}))\cdot \textrm{min}\{\epsilon t,1\},
\ee
\be \nonumber
\im\int_{\varphi^{*}_{br-}}^{\varphi_{br-}}\frac{\varepsilon(\varphi)}{E(\varphi)}e^{itE(\varphi)}d\varphi=\gamma^{-1}\epsilon+O(\epsilon^{2}\gamma^{-2}).
\ee
Combining these results with contributions from saddle points $\varphi_{2}$, $\varphi_{3}$, $\varphi_{4},$ one obtains
\be \label{asymptotic A_0 h=k}
A_{0}(t) \simeq \sqrt{\frac{1}{2 \pi(2-\gamma^{2}) t}}\left(\sqrt{2}\cos(2t-\frac{\pi}{4}) \right.
\ee
\be \nonumber
\left.+(1-\gamma^{2})^{\frac{1}{4}} \exp[-\frac{\gamma^{2}}{\sqrt{1-\gamma^{2}}} t] +O(\frac{1}{t})+O(\epsilon \gamma^{-1})\right).
\ee
For $B_{0}$, keeping in mind (\ref{saddle points for B_0}), one obtains
\be \label{asymptotic B_0 h=k}
B_{0}(t) \simeq \sqrt{\frac{1}{2 \pi(2-\gamma^{2}) t}}\left(\sqrt{2} \sin(2t-\frac{\pi}{4}) \right.
\ee
\be \nonumber
\left.+ (1-\gamma^{2})^{-\frac{1}{4}} \exp[-\frac{\gamma^{2}}{\sqrt{1-\gamma^{2}}} t] +O(\frac{1}{t})+O(\epsilon \gamma^{-1})\right).
\ee
Region of applicability for these asymptotics is limited by the condition that $\exp[-\frac{1}{2}|E''(\varphi_{3})|R^{2}]$ must be small enough (here R is the radius of convergence for $E(\varphi)$ near $\varphi_{3,4}$). We have $R=|\varphi_{3}-\varphi_{br+}|\simeq (2-\sqrt{2})\gamma$ for small $\gamma$. Thus we get the following condition of applicability for the above asymptotic:
\be
t > (2-\sqrt{2})^{-2}\gamma^{-2}\ln(\frac{1}{\vartriangle}),
\ee
where  $\vartriangle$ is the order of the relative error.
Under this condition we can neglect the term $\exp[-\gamma^{2}(1-\gamma^{2})^{-\frac{1}{2}} t]$ in (\ref{asymptotic A_0 h=k}) and (\ref{asymptotic B_0 h=k}). Numerical evaluation shows excellent coincidence of these asymptotics with the exact values of $A_{0}$ and $B_{0}$ in the case $\gamma \ll 1,~ t \gg 1$. Moreover, these expressions exactly coincide with asymptotic forms for Bessel functions in (\ref{A and B through Bessel}) in the case $\gamma=0,~ h=1$ which is not obvious from our derivation method. With these remarques, we find

$$
g^{zz}_{0}(t)= \frac{1}{ \pi(2-\gamma^2)t}
\left(
1+\frac{2-\gamma^2}{2\sqrt{1-\gamma^2}}\exp[-\frac{2\gamma^{2}}{\sqrt{1-\gamma^{2}}} t]
\right.
$$
\be
\left.
+2\sqrt{\frac{2-\gamma^2}{2\sqrt{(1-\gamma^2)}}}\exp[-\frac{\gamma^{2}}{\sqrt{1-\gamma^{2}}} t] \cos(2t-\frac{\pi}{4}-\arctan\frac1{\sqrt{1-\gamma^2}})
\right),
\ee
which gives us an excellent approximation for $g^{zz}_{0}(t)$ in the case $\epsilon \ll \gamma,~\epsilon \ll N^{-1},~ t<t_{\rm th}$. For not very small $\gamma^{2}$ we can neglect exponential suppressed terms and obtain
\be
g^{zz}_{0}(t)= \frac{1}{ \pi(2-\gamma^2)t}.
\ee

\subsection{Asymptotics for spectral functions of non-zero order}

For spectral functions of non-zero orders the situation is slightly more complicated.  Again there are four saddle points in the strip $\re(\varphi)\in(-\pi,\pi]$, but now their positions {\it vary with time} (see fig. \ref{asyptotic non-zero order}). Consider the function $A_{j},$  $j\geq1$. The corresponding saddle points satisfy the 4-degree polynomial equation on $z= \cos\varphi:$
\be \label{eq on E'(z)}
\begin{array}{lcl}
 (1-\gamma ^{2})^{2}z^{4}-2h(1-\gamma ^{2})z^{3}+[((h^{2}-(1-\gamma ^{2})^2))+\zeta (1-\gamma^{2})]z^{2}+\\
+[h(1-\gamma ^{2})-2\zeta h]z- h^{2}+\zeta (h^{2}+\gamma ^{2})=0,
\end{array}
\ee
where $\zeta\equiv (Nj)^{2}(t)^{-2}$. This equation viewed as the equation on $\varphi$ gives eight solutions in the strip $\re(\varphi)\in(-\pi,\pi]$. Four of them are relevant (i.e. are solutions of eq. (\ref{general eq for saddle points})) and other four are irrelevant (are solutions of the equation $-E't-jN=0$).
Since $E'(\varphi)$ takes all possible real values on the each branch cut, one pair of saddle points, $\varphi_{3}$ and $\varphi_{4}$, lies on two brunch cuts symmetrically with respect to the real axis, analogous to the $j=0$ case.

Let us consider the positions of two other saddle points, $\varphi_{1}$ and $\varphi_{2}$, especially their evolution with time. The definition of the threshold time implies that for $t<j t_{\rm th}$ eq.(\ref{general eq for saddle points}) has no real roots. Thus $\varphi_{1}$ and $\varphi_{2}$ are complex. In fact they are complex conjugate to each other. When time goes on they both approach $\varphi_0$ which lies on the real axis, and eventually merge at $t=jt_{\rm th}:$ $\varphi_1(jt_{\rm th})=\varphi_2(jt_{\rm th})=\varphi_0.$ For $t>j t_{\rm th}$ $\varphi_{1}$ and $\varphi_{2}$ lie on the real axis and move apart from $\varphi_0$ and from each other, approaching $0$ and $\pi$ correspondingly as $t\rightarrow\infty.$

\subsubsection{\label{appendix t>jtTh}Asymptotics for $t>jt_{\rm th}$}


To start with, we warn the reader that we do not provide a strict mathematical proof that the suitable integration path exists which goes through the chosen saddle points in all presented cases. However, the existing of these paths looks quite natural in all cases and, moreover, corresponding asymptotic expressions show excellent coincidence with numerical evolutions. Strict mathematical proof is postponed for further work.

With this warning made, let us turn to the case $t>jt_{\rm th}$. We start from the case of non-small $\epsilon$ and we assume that $\varphi_{1}$ and $\varphi_{2}$ are situated far enough from each other, so that we can neglect their mutual influence in the asymptotics. The conditions under which this assumption is fulfilled are considered in what follows.

\begin{figure}
\center{
\begin{tabular}{lr}
\includegraphics[width=0.5\textwidth]{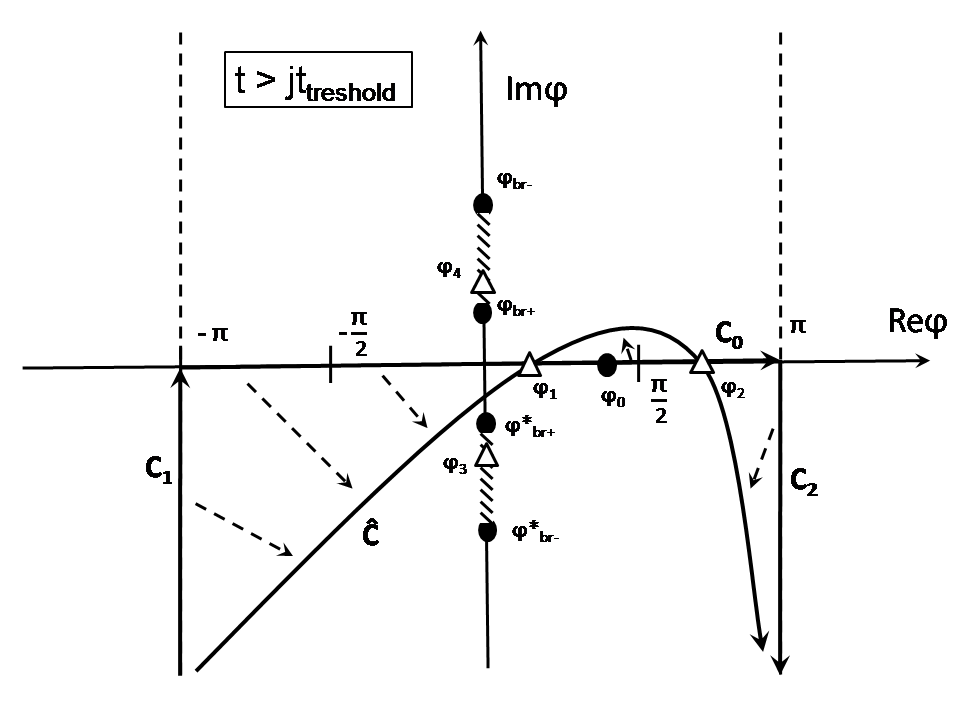} &
\includegraphics[width=0.5\textwidth]{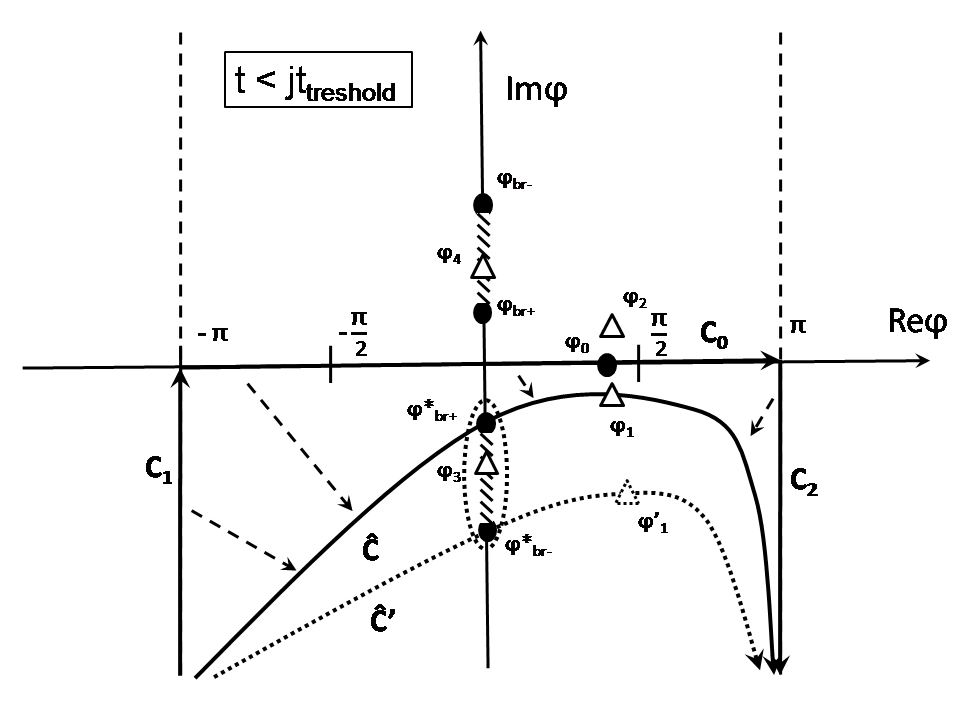}
\end{tabular}
}
\caption{\label{asyptotic non-zero order} Integration paths for $A_{j}$, $B_{j}$ in the case $t>jt_{\rm th}$ (left) and in the case $t<jt_{\rm th}$ (right). Empty triangles correspond to saddle points, filled circles -- to branch points and to the point $\varphi_{0}$ which satisfies $E(\varphi_{0})''=0$.}
\end{figure}

Note that we can integrate along the path $C_{1}\cup C_{0}\cup C_{2}$, where $C_{0}$ is the original path,  $C_{1}$ starts from $\varphi=-\pi-i\infty$ and goes to $\varphi=-\pi$ along $\re \varphi=\textrm{const}=-\pi$, $C_{2}$ starts from $\varphi=\pi$  and goes to $\varphi=\pi-i\infty$ along $\re \varphi=\textrm{const}=\pi$ (see fig. \ref{asyptotic non-zero order}). Since $f_{A,B j}(x+iy)$ are $2\pi$ periodic functions of $x$, the value of the integral along the new path is exactly the same as along $C_{0}$. Now we transform the path  $C_{1}\cup C_{0}\cup C_{2}$ to the path $\widehat{C}$ which starts from $\varphi=-\pi - i\infty$, goes 
through saddle points $\varphi_{1}$ and $\varphi_{2}$ and ends at $\varphi=\pi-i\infty$ (see fig. \ref{asyptotic non-zero order}, left). There are two topological different possible integration paths $\widehat{C}$: the first one goes above the branch cut, while the second one - under the cut. In the latter case according to the Cauchy theorem  we have to subtract
the integral over the branch cut. However as was discussed above this integral does not contribute to $A_j$ and $B_j$ (see eq.(\ref{integration through incision})).

Now we can proceed to find asymptotic expressions as contribution from points $\varphi_{1}$ and $\varphi_{2}$. Under all specified conditions, we immediately obtain eq. (\ref{formula for A asymptotic non-zero order}) for $A_j.$
For $B_{j}$, using reasonings similar to those  following  eq. (\ref{saddle points for B_0}), we get eq. (\ref{formula for B asymptotic non-zero order}).

Let us now investigate the range of applicability of eqs. (\ref{formula for A asymptotic non-zero order}), (\ref{formula for B asymptotic non-zero order}).
Firstly we consider in what cases we can use the standard approximation for contribution of saddle points under the assumption that radius of convergence for corresponding series is large enough. In this case the derived approximation  may deviate from the exact expression for two reasons: small value of $|E(\varphi_{1,2})''|$ and interception of contributions for $\varphi_{1}$ and $\varphi_{2}$ due to their close relative position. These two features can appear only for small times after $jt_{\rm th}$. Let us give more precise estimation without detail explanations. If $\delta t \equiv t-jt_{\rm th}>0$, then it has to be $\frac{\delta t}{t_{\rm th}}>\sqrt[3]{\frac{j}{2 \cdot r \cdot N^{2}}}$, where $r \equiv \frac{E(\varphi_{0})}{|E'''(\varphi_{0})|}$ is quantity of order of $1$ for the vast majority of the Hamiltonian parameter space. One can see that these asymptotic approximations for spectral functions of order $j$ become accurate starting from time close to $j t_{\rm th}.$

Now let us investigate in what cases series does not converge in large enough circle for some saddle point. We have to explore small enough $\epsilon$, at least $\epsilon^{-1} \gg t.$  Since for spectral function of order $j$ we are interested in $t>j t_{\rm th}$, it is useful to consider $\epsilon^{-1} \gg j  N$. Firstly we define the position of $\varphi_{0}:$
\be
\varphi_{0}=\arccos(\frac{2\gamma^{2}+1-\sqrt{4\gamma^{2}+1}}{2(1-\gamma^{2})})+O(\epsilon),~~~\gamma^{2}\leq\frac{3}{4},
\ee
\be \nonumber
\varphi_{0}=\frac{2^{\frac{1}{2}}\epsilon^{\frac{1}{2}}}{(4\gamma^{2}-3)^{\frac{1}{4}}}+O(\epsilon),~~~~~~~~~\gamma^{2}>\frac{3}{4}.
\ee
Let us consider the case $\gamma^{2}<\frac{3}{4}$. For times which satisfy $jt_{\rm th}<t<j\widetilde{t}_{\rm th}=jN\gamma^{-1}+O(\epsilon)$  there is no point $\varphi_{1}$ near $\varphi=0$, and the derived asymptotic expressions (\ref{formula for A asymptotic non-zero order}), (\ref{formula for B asymptotic non-zero order}) are valid. If $t>\widetilde{t}_{\rm th}$ one obtains
\be
\varphi_{1}=\frac{\epsilon}{\gamma\sqrt{\frac{\gamma^{2}t^{2}}{j^{2}N^{2}}-1}}+O(\epsilon^{2}),~~~t>j\widetilde{t}_{\rm th}=j\frac{N}{\gamma}+O(\epsilon).
\ee
Thus for $t \ll \epsilon^{-1}$ we cannot use the above derived approximations because $\varphi_{1}$ is situated close to $\varphi_{br-}$ and the radius of convergence $R=\sqrt{\epsilon^{2}\gamma^{2}+\varphi_{1}^{2}}$ is very small,  $|E''(\varphi_{1})|tR^{2} < 1$, thus we cannot use the method of the steepest descent for $\varphi_{1}$ (we assume here $\gamma \gg \epsilon$). Instead we can proceed analogously to the case of spectral functions of zero order. Namely, we move the integration path in order to go through saddle points $\varphi_{3}$ and $\varphi_{4}$ and neglect the value of integral between $\varphi^{*}_{br-}$ and $\varphi_{br-}$ as we have done in (\ref{zero order spectral function asymptotic}). The difference from the case of $A_{0}$ and $B_{0}$ is that now we can neglect exponentially suppressed contribution from $\varphi_{3}$, but contribution from $\varphi_{4}$ may be not  small for some portion of time. Numerical evaluation shows that for $N\gg1$ the contribution from $\varphi_{4}$ is exponentially suppressed at a timescale $\sim \widetilde{t}_{\rm th}.$
 Summarizing, for time $t>\widetilde{t}_{\rm th}$ we obtain
\be \label{asympA h=1 fi1 near 0}
A_j(t)\simeq \frac{1}{\sqrt{2\pi t}}\left(\frac{1}{2}\sqrt{\frac{1}{|E''(\varphi_4)|}}\exp[-t(-i E(\varphi_4+0))+jN(-i\varphi_4)]\right.
\ee
\be \nonumber
+\left.\sqrt{\frac{1}{|E''(\varphi_2)|}}\cos( tE(\varphi_2)-jN\varphi_2-\frac{\pi}{4})+O(\frac{1}{t})\right),
\ee
where one should remind that
$-iE(\varphi_4+0)>0,~(-i\varphi_4)>0.$  Factor $\frac{1}{2}$ in due to the fact that  we have to take only one half of contribution from $\varphi_{4}$. Analogously one obtains
\be \label{asympB h=1 fi1 near 0}
B_j(t)\simeq \frac{1}{\sqrt{2\pi t}}\left(\frac{1}{2}\frac{ \varepsilon(\varphi_{4})}{(-iE(\varphi_{4}+0))}\sqrt{\frac{1}{|E''(\varphi_4)|}}\exp[-t(-iE(\varphi_4+0))+jN(-i\varphi_4)]\right.
\ee
\be \nonumber
+\left.\frac{ \varepsilon(\varphi_{2})}{E(\varphi_{2})}\sqrt{\frac{1}{|E''(\varphi_2)|}}\sin( tE(\varphi_2)-jN\varphi_2-\frac{\pi}{4})+O(\frac{1}{t})\right).
\ee
The only case we do not investigate here is $\gamma \leq \epsilon \ll 1.$ We  leave it for future work.

If $\gamma^{2} \geq \frac{3}{4}$, then $\varphi_{0}$ is situated in the vicinity of $\varphi=0$, which leads to $t_{\rm th}=\widetilde{t}_{\rm th}=N\gamma^{-1}+O(\epsilon)$. Therefore asymptotic expressions (\ref{asympA h=1 fi1 near 0}), (\ref{asympB h=1 fi1 near 0}) are valid starting from time t for which $\varphi_{2}$ is situated far enough from $\varphi=0$ (the corresponding condition reads $\frac{1}{2}|E''(\varphi_2)|t\varphi_{2}^{2} \gg 1$). For times in the vicinity of  $j t_{\rm th}$ another approximation should be used, see Sec. \ref{appendix time near jt_th}.

\subsubsection{\label{appendix t<jtTh}Asymptotics for $t<jt_{\rm th}$}

When $t<jt_{\rm th}$ the path of integration should not go through the I quadrant, since for all values of $x\in(-\pi;\pi)$ $\partial_{y}\re (iEt - in\varphi)>0$ (remind that $\varphi = x + i y$). We shift our integration path to $\widetilde{C}$ or $\widetilde{C}'$ (see figure (\ref{asyptotic non-zero order})) and obtain asymptotics from the contribution from only one saddle point, $\varphi_{1}$. For large enough $|\delta t|$, where $\delta t \equiv t-jt_{\rm th}<0$, we have
\be \label{exact suppression for large delta t}
A_{j}(t)\simeq \sqrt{\frac{1}{2\pi |E''(\varphi_{1})|t}}\cos( E(\varphi_{1})t-jN\varphi_{1}+\sigma(\varphi_{1})),
\ee
where $\sigma(\varphi_{1})$ is the angle between the path and the line $\im f_{A_j}={\rm const}.$
Let us evaluate $\varphi_{1}$ approximately for  $t\lesssim jt_{\rm th}$ such that $t^{-1}\delta t  \ll 1$. If $E'''_{0} \equiv E'''(\varphi_{0})\neq 0$, then
\be \label{approximate fi suppression case}
\varphi_{1}=\varphi_{0}-i\eta+\frac{1}{6}\frac{E^{(IV)}_{0}}{E^{'''}_{0}}\eta^{2}+i(\frac{5}{72}(\frac{E^{(IV)}_{0}}{E^{'''}_{0}})^{2}-\frac{1}{24}E^{(V)}_{0})\eta^{3}+O(\eta^{4})+iO(\eta^{5}),
\ee
where
\be \nonumber
 \eta \equiv \sqrt{-\frac{2N\delta t}{|E^{'''}_{0}|tt_{\rm th}}}=\sqrt{-2r{\frac{\delta t}{t}}},~~r\equiv\frac{E'_{0}}{|E^{'''}_{0}|}
\ee
Up to the  first order in $\delta t \cdot t^{-1} \simeq \delta t \cdot j^{-1} t_{\rm th}^{-1}$ one obtains

\be \label{approximate suppression}
A_{j}(t) \simeq \frac{r^{\frac{1}{4}}(jN)^{\frac{1}{2}}}{2^{\frac{3}{4}}\pi^{\frac{1}{2}}}(\frac{-\delta t}{jt_{\rm th}})^{-\frac{1}{4}}
 \exp[\frac{-2\sqrt{2r}}{3} jN  (\frac{-\delta t}{jt_{\rm th}})^{\frac{3}{2}}+O((\frac{-\delta t}{jt_{\rm th}})^{\frac{5}{2}})] \cos(E(\varphi_{0})t-jN\varphi_{0})
\ee
\be \nonumber
+O((\frac{-\delta t}{jt_{\rm th}})^{\frac{3}{4}}).
\ee
This expressions gives  only the order of suppression; if one is interesting in more precise expression, he has to directly solve eq.(\ref{general eq for saddle points}) to find an exact value of $\varphi_1$  and substitute it in the  general formula (\ref{exact suppression for large delta t}).
This formula (and approximation (\ref{approximate suppression})) is valid until we can neglect the term $\frac{1}{6}E'''(\varphi_{1})(\vartriangle \varphi)^{3}$ in comparison with $\frac{1}{2}E''(\varphi_{1})(\vartriangle \varphi)^{2}$ in series expansion near $\varphi_{1}$. This leads to the condition $jN(\frac{\delta t}{jt_{\rm th}})^{\frac{3}{2}} \gg 1$. In the opposite case,  $jN(\frac{\delta t}{jt_{\rm th}})^{\frac{3}{2}} \ll 1$, eqs. (\ref{general eq for saddle points}) and (\ref{approximate suppression}) are not valid and the leading order contribution is given by the term $\frac{1}{6}E'''(\varphi_{1})(\vartriangle \varphi)^{3}$, which leads to


\be \label{approximate suppression near transition time}
A_{j}(t) \simeq \frac{\Gamma(\frac{1}{3})r^{\frac{1}{3}}}{2^\frac{2}{3}3^{\frac{1}{6}}\pi j^{\frac{1}{3}}N^{\frac{1}{3}}}\cdot \left(\frac{|E(\varphi_{0})'''|t_{\rm th}}{|E(\varphi_{1}''')|t}\right)^{\frac{1}{3}}\cdot \cos(E(\varphi_{1})t-jN\varphi_{1}),
\ee
where values of $E(\varphi_{1})$ and $E(\varphi_{1}''')$ may be found according to eq. (\ref{approximate fi suppression case}).
Eqs. (\ref{approximate suppression near transition time}) and (\ref{approximate suppression}) become asymptotics for $J_{j}(t)$ in the case of  $XX$ chain. If one wants to derive asymptotics  valid in the region $jN(\frac{\delta t}{jt_{\rm th}})^{\frac{3}{2}}\sim 1$, he has to calculate the integral through saddle point with proper path direction (here we use approximation (\ref{approximate fi suppression case}) for saddle points),
\be \label{integral near saddle point}
I_{\textrm{saddle}}(t)\simeq\int \exp[-\frac{1}{2}\eta|E'''_{0}|tz^{2}-\frac{i}{6}|E'''_{0}|tz^{3}]dz,
\ee
and use the formula
\be
A_{j}(t)\simeq \frac{1}{2\pi}I_{\textrm{saddle}}(t)\exp[\frac{-2\sqrt{2r}}{3} jN  (\frac{-\delta t}{jt_{\rm th}})^{\frac{3}{2}}]\cos(E(\varphi_{0})t-jN\varphi_{0}).
\ee
Eqs. (\ref{approximate suppression}) and (\ref{approximate suppression near transition time}) can be obtained from the above formula by neglecting the second summand in the exponent in (\ref{integral near saddle point}) and by  expansion of the integrand in eq.(\ref{integral near saddle point}) in powers of $\eta$.

Let us turn to $B_{j}.$ In order to describe its behavior in analogous way one should start from
\be \label{suppression for B}
B_{j}(t)\simeq \Im \left(\sqrt{\frac{1}{2\pi |E''(\varphi_{1})|t}}\frac{\varepsilon(\varphi_{1})}{E(\varphi_{1})}\exp[ iE(\varphi_{1})t-ijN\varphi_{1}+i\sigma(\varphi_{1})]\right)
\ee
instead of (\ref{exact suppression for large delta t}). We do not describe in details $B_{j}(t)$
because there is no simple approximation formula for (\ref{suppression for B}) for all possible values of $\cos(\varphi_{0}).$
However the exponential suppression for $B_{j}(t)$ has the same form as for $A_{j}(t)$, only the preexponential factor differs. This is because the suppression is determined by the exponent of the quantity $-\Re(-iE(\varphi_{1})t+ijN\varphi_{1})$ which is the same for $B_{j}$ and $A_{j}$ up to $t^{-1}$.

To conclude this subsection,  the suppression of spectral functions $A_{j}(t)$ and $B_{j}(t)$ with $\delta t = t - j t_{\rm th}<0$ with exponential precision reads
\be \label{low of suppression}
A_{j}(t)\sim B_{j}(t)\sim \exp[-\frac{2}{3}\sqrt{\frac{2E'(\varphi_{0})}{|E'''(\varphi_{0})|}} jN  (\frac{-\delta t}{jt_{\rm th}})^{\frac{3}{2}}+O((\frac{-\delta t}{jt_{\rm th}})^{\frac{5}{2}})].
\ee
This is in accordance with a general result \cite{lieb1972finite}.
Here we assume that $E'''(\varphi_{0}) \neq 0$; for very small $E'''(\varphi_{0})$ one has to take into account $E^{(IV)}(\varphi_{0})$ which leads to a slightly different law.

\subsubsection{\label{appendix time near jt_th}Asymptotics for $t \simeq jt_{\rm th}$ }

The considerations of the present subsection are less rigorous than in the previous ones. The only consequence, however, is that we do not completely control errors for derived approximations. Numerical calculations demonstrate that the latter are nevertheless rather accurate  in a wide range of model parameters. For certain regions of the parameter space in which the derived  approximations fail we are able to identify the reason and point out the way to overcome the difficulties.

Let us describe the method. When $t \simeq t_{\rm th}$ the saddle points are situated near $\varphi_{0}$. Integrand is a very fast oscillating function for all $\varphi$ except $\varphi \simeq \varphi_{0}.$ Thus the value of the integral is picked up on a small segment [$\varphi_{0}- \vartriangle \varphi ; \varphi_{0}+ \vartriangle \varphi  $], $\vartriangle \varphi > \frac{1}{2}(\varphi_{2}-\varphi_{1})$. 
In order to estimate errors for this approximation, one has to make bulky calculations in the spirit of the above subsections. We avoid this in the present work.

In order to calculate the integral along the small segment, we expand $E(\varphi)$ in the vicinity of $\varphi_{0}.$  And the last approximation is to replace the interval of integration from $[- \vartriangle \varphi, \vartriangle \varphi]$ to $[-\infty; \infty],$ where the integration variable is $\mu\equiv\varphi - \varphi_{0}.$ The latter trick is justified because our new integrand oscillate as $\sim \exp(i\cdot \textrm{const} \cdot \mu^{3})$ when $\mu \rightarrow \pm \infty$. All these approximations are legitimate when the model parameters are such that $\varphi_{0}$ is far enough from points of branching. Thus the approximation is valid for large enough $\epsilon$ and all values of $\gamma$ and for $\epsilon \ll 1$ it is valid for $\gamma^{2} < \frac{3}{4}.$ Let us assume that $E_{0}''' \equiv E'''(\varphi_{0})$ is not very small. In this case one can consider power expansion of $E(\varphi)$ up to $(\varphi-\varphi_{0})^{3}$ and neglect terms $\sim O((\varphi-\varphi_{0})^{4})$. Thus for $t \simeq jt_{\rm th}$ one gets
\be \nonumber \label{Airy deriviation}
A_{j}(t)=\frac{1}{2\pi}\int_{-\pi}^{\pi}\cos(Et-jN\varphi)d\varphi \simeq \frac{1}{2\pi}\int_{-\vartriangle \varphi}^{\vartriangle \varphi}\cos(E_{0}t-jN\varphi_{0}+(E_{0}'t-jN)\mu-\frac{1}{6}|E_{0}'''|t\mu^{3})  d\mu
\ee
\be
\simeq \left(\frac{2}{|E_{0}'''|t} \right)^{\frac{1}{3}}\frac{1}{2\pi}\int_{-\infty}^{\infty}\cos[(E_{0}t-jN\varphi_{0})-\frac{t-jt_{\rm th}}{t_{\rm th}}N\left(\frac{2}{|E_{0}'''|t} \right)^{\frac{1}{3}}\xi+\frac{1}{3}\xi^{3}]  d\xi.
\ee
Here all functions $f$ with subindex $0$ should be understood as $f_{0}\equiv f(\varphi_{0})$.
The above integral is a well-known Airy function of the first kind:
\be
\textrm{Ai}(x) \equiv \frac{1}{\pi}\int_{0}^{\infty}\cos(\frac{\xi^{3}}{3}+x\xi)d\xi
\ee
Thus we get eq. (\ref{airy  Aj}) (with analogously reasonings for $B_{j}$).

Let us investigate now the case $h=1$, $\gamma^{2}>\frac{3}{4}$ (we do not investigate here the case of small, but non-zero $\epsilon=h-1 \ll 1$). For $h=1,~\gamma^{2}>\frac{3}{4}$ we cannot use formula (\ref{Airy deriviation}), since $\varphi_{0}=\varphi_{br-}=\varphi_{br-}^{*}=0$. 
But we can argue that the main contribution for $A_{j}$ is picked up on the segment $[0, \vartriangle \varphi]$, since only on this segment oscillation frequency is not very large (note that for $[-\vartriangle \varphi, 0]$ the frequency is large; this is due to the discontinuity of the group velocity at $\varphi=0$). If we somehow expand $E(\varphi)$ on $[0, \vartriangle \varphi]$, we can obtain a good approximation. Consider -$E(\varphi)$ on a complex plain. If we have values  for $E(\varphi)$ on the segment $\varphi \in [0,\pi]$ fixed, we can make analytical continuation to $\Re (\varphi) \in [-\pi,0]$ in two different ways: with the branch cut: $\Re (\varphi_{\textrm{branch cut}}) =0 ,~\Im (\varphi_{\textrm{branch cut}}) \in [-\varphi_{br+}^{*},\varphi_{br+}]$ or with the branch cut: $\Re (\varphi_{\textrm{branch cut}}) =0 ,\Im (\varphi_{\textrm{branch cut}}) \in [-\infty,-\varphi_{br+}^{*}]\cup[\varphi_{br+},\infty]$. In the first case we have our original function $E(\varphi)$ in the segment $\varphi \in [-\pi,0]$. In the second case we have some new function $\widetilde{E}(\varphi)$ which does not coincide with $E(\varphi)$ in this segment. But in the latter case we can use power expansion for $\widetilde{E}(\varphi)$ in the circle with radius $R=|\varphi_{br+}|$:
\be \label{E tidle approx}
\widetilde{E}(\varphi)=\gamma \varphi+\frac{3-4\gamma^{2}}{24\gamma}\varphi^{3}+\frac{16\gamma^{4}-15}{1920\gamma^{3}}\varphi^{5}+O(\varphi^{7}).
\ee
Analogously to (\ref{Airy deriviation}) one obtains
\be \label{airy general integral for h=1 gamma2 near 075}
A_{j}(t)\simeq \frac{1}{2\pi}\int_{0}^{\infty}\cos[(\gamma t - jN) \varphi+\frac{3-4\gamma^{2}}{24\gamma}t\varphi^{3}+\frac{16\gamma^{4}-15}{1920\gamma^{3}}t\varphi^{5}]d\varphi.
\ee
If one can neglect the term $\sim \varphi^{5}$ (i.e. if $\gamma^{2}-\frac{3}{4}$ is large enough),  he gets
\be \label{airy h=1 gamma large app Aj}
A_{j}(t) \simeq \frac{1}{2}\left(\frac{2\gamma}{(\gamma^{2}-\frac{3}{4})t} \right)^{\frac{1}{3}}\cdot \textrm{Ai} \left[-(\gamma t -jN)\left(\frac{2\gamma}{(\gamma^{2}-\frac{3}{4})t} \right)^{\frac{1}{3}} \right],~~~h=1,~\gamma^{2}>\frac{3}{4}.
\ee
For $B_{j}$ one uses  power expansion $\varepsilon(\varphi)E^{-1}(\varphi)$ in the vicinity of $\varphi = 0$ to obtain
\be \label{airy h=1 gamma large app Bj}
B_{j}(t) \simeq \frac{1}{4\gamma}\left(\frac{2\gamma}{(\gamma^{2}-\frac{3}{4})t} \right)^{\frac{2}{3}}\cdot \textrm{Ai}' \left[-(\gamma t -jN)\left(\frac{2\gamma}{(\gamma^{2}-\frac{3}{4})t} \right)^{\frac{1}{3}} \right],~~~h=1,~\gamma^{2}>\frac{3}{4},
\ee
where prime stands for the derivative.

Let us consider the special case $h=1,~\gamma^{2}=\frac{3}{4}$. We have $t_{\rm th}=\frac{2}{\sqrt{3}}N$. The term $ \sim \varphi^{3}$ in power expansion $\widetilde{E}(\varphi)$ is zero: $\widetilde{E}(\varphi)=\frac{\sqrt{3}}{2}\varphi-\frac{1}{120\sqrt{3}}\varphi^{5}$. Let us introduce the function
\be
\textrm{gAi}_{n}(x)\equiv \frac{1}{\pi}\int_{0}^{\infty}\cos(\frac{\xi^{n}}{n}+x\xi)d\xi
\ee
Airy function of the first kind is a particular case of this function: $\textrm{Ai}(x)=\textrm{gAi}_{3}(x)$. $\textrm{gAi}_{n}(x)$ for $n>3$ exhibits  the same behavior as $\textrm{Ai}(x)$: for positive $x$ it is exponentially decreasing, and for negative $x$ it oscillates and goes to zero when $x\rightarrow -\infty.$ $A_{j}$ and $B_{j}$ are expressed through this function and its derivative:
\be \label{airy h=1 gamma = 075 app Aj}
A_{j}(t) \simeq \frac{1}{2}\left(\frac{24\sqrt{3}}{t} \right)^{\frac{1}{5}} \cdot \textrm{gAi}_{5}\left[-(\frac{\sqrt{3}t}{2}-jN)\left(\frac{24\sqrt{3}}{t} \right)^{\frac{1}{5}}\right],~~~h=1,~\gamma^{2}=\frac{3}{4}
\ee
\be \nonumber
B_{j}(t) \simeq \frac{1}{4\gamma}\left(\frac{24\sqrt{3}}{t} \right)^{\frac{2}{5}} \cdot \textrm{gAi}_{5}'\left[-(\frac{\sqrt{3}t}{2}-jN)\left(\frac{24\sqrt{3}}{t} \right)^{\frac{1}{5}}\right],~~~h=1,~\gamma^{2}=\frac{3}{4}
\ee
Note that rather small value of the coefficient of term $\varphi^{5}$ in power expansion of $\widetilde{E}(\varphi)$ implies that these approximations work well for sufficiently large time ($t \gg 24\sqrt{3}$). Analogously we require $t \gg 2\gamma(\gamma^{2}-\frac{3}{4})^{-1}$ in eqs. (\ref{airy h=1 gamma large app Aj}), (\ref{airy h=1 gamma large app Bj}). Note that if one wishes to investigate approximations which successfully describe $A_{j}$, $B_{j}$ near threshold time in the region of parameter space $h=1, \gamma^{2} \simeq \frac{3}{4}$, then one has to calculate integral (\ref{airy general integral for h=1 gamma2 near 075}) saving both terms $\sim \varphi^{3}$ and $\sim \varphi^{5}$. Therefore, in this case there is no such clear power law for time dependence of maximum value of spectral function as in (\ref{maximums for Aj}).

Now let us discuss possible values of maximums for $A_{j}$ and $B_{j}$. If $j N$ is sufficiently large the positions of global maximums of $A_{j}$ and $B_{j}$  coincide with those for $\textrm{Ai}(x)$ and $\textrm{gAi}_{5}(x):$
\be \label{maximums for Aj}
\sup\limits_t A_{j}(t) \simeq \left(\frac{2}{|E'''_{0}|t}\right)^{\frac{1}{3}} \cdot a_{3}, ~~~\textrm{for} ~h>1;~\textrm{or}~ h=1, \gamma^{2} < \frac{3}{4}
\ee
\be \nonumber
\sup\limits_t A_{j}(t) \simeq \frac{1}{2}\left(\frac{2\gamma}{(\gamma^{2}-\frac{3}{4})t} \right)^{\frac{1}{3}} \cdot a_{3},~~~\textrm{for}~h=1, \gamma^{2}>\frac{3}{4}
\ee
\be \nonumber
\sup\limits_t A_{j}(t) \simeq \frac{1}{2}\left(\frac{24\sqrt{3}}{t} \right)^{\frac{1}{5}} \cdot a_{5},~~~\textrm{for}~h=1, \gamma^{2}=\frac{3}{4}
\ee
Here $a_{3}$ and $a_{5}$ are global maximums of $\textrm{Ai}(x)$ and $\textrm{gAi}_{5}(x)~$ ($a_{3}=0,54...$, $a_{5}=0.44...$) correspondingly.  Without loss of precision one can replace $t$  by $jt_{\rm th}$ in the above expressions. We do not present the analogous expressions for $B_{j}(t)$ because positions of maximums of these functions obviously do not coincide with those for $A_{j}$ (when $A_{j}(t)$ achieves global maximum $B_{j}(t)$ becomes zero).  In the last two cases in eq. (\ref{maximums for Aj}) it is $A_{j}^{2}$ which gives the  main contribution in $g_{0}^{zz}(t)$ near $t=jt_{\rm th}.$
When  $h>1$ or $h=1, \gamma^{2}<\frac{3}{4}$ one has to  investigate maximum of $A_{j}(t)^{2}+ B_{j}(t)^{2}$ in order to find the leading contribution to $g_{0}^{zz}(t).$  This leads to
\be\label{revival maxima}
\begin{array}{lr}
\textrm{Max}_{(\textrm{revival j})}[g^{zz}_{0}(t)] \simeq 4\left(\frac{2}{|E_{0}'''|j t_{\rm th}} \right)^{\frac{2}{3}}\cdot(1+|\frac{\varepsilon_{0}}{E_{0}}-1|) a_{3}^{2} & \textrm{for} ~h>1;~\textrm{or}~ h=1, \gamma^{2} < \frac{3}{4},\\
\textrm{Max}_{(\textrm{revival j})}[g^{zz}_{0}(t)] \simeq \left(\frac{2\gamma^{2}}{(\gamma^{2}-\frac{3}{4})jN} \right)^{\frac{2}{3}} \cdot a_{3}^{2}, & \textrm{for} ~ h=1, \gamma^{2} > \frac{3}{4},\\
\textrm{Max}_{(\textrm{revival j})}[g^{zz}_{0}(t)] \simeq \left(\frac{36}{jN} \right)^{\frac{2}{5}}\cdot a_{5}^{2}, & \textrm{for}~h=1, \gamma^{2}=\frac{3}{4}.
\end{array}
\ee

Note that this is a rather rude approximation, since interference terms may be large. Therefore these expressions work well only for large number of spins. Numerical evolution shows that the revivals are maximally pronounced for $h=1$ and $\gamma^{2}$ slightly less then $\frac{3}{4}.$   This may be expected on the basis of eq. (\ref{airy general integral for h=1 gamma2 near 075}).

Let us emphasize once more time  that all the derived expressions can be derived with more rigor using integrals in the complex plain analogously to that was done in the previous subsections.
We have not explored the whole parameter space. In particular, we have not considered the cases $h=1, \gamma^{2} \simeq \frac{3}{4}$ , $\epsilon=h-1 \ll 1, \epsilon \neq 0$, or $h>1, E_{0}'''=0$. Some of the derived approximations work well only for large time and, correspondingly, large number of spins (for example, $t \gg 24\sqrt{3}$ or $t \gg 2\gamma(\gamma^{2}-\frac{3}{4})^{-1}$). However, for a large region of parameter space these approximations work fairly well, and they provide an opportunity to investigate amplitudes of maximums in partial revivals, or at least the law of there decrease. For these reasons we decided to include in the paper  these not completely rigorous calculations. The validity of formulae derived in the present  subsection is justified by the fact that  approximations (\ref{airy  Aj}) give the same result as the more rigorously derived eqs. (\ref{approximate suppression}) and (\ref{approximate suppression near transition time}) when the ranges of applicability overlap.


\bibliographystyle{unsrt}	
\bibliography{spin_chains}	

\begin{thebibliography}{10}

\bibitem{PorrasCirac}
D.~Porras and J.~I. Cirac.
\newblock Effective quantum spin systems with trapped ions.
\newblock {\em Phys. Rev. Lett.}, 92(20):207901, May 2004.

\bibitem{bose2003quantum}
S.~Bose.
\newblock Quantum communication through an unmodulated spin chain.
\newblock {\em Physical review letters}, 91(20):207901, 2003.

\bibitem{lieb1961two}
E.~Lieb, T.~Schultz, and D.~Mattis.
\newblock Two soluble models of an antiferromagnetic chain.
\newblock {\em Annals of Physics}, 16(3):407--466, 1961.

\bibitem{mazur1973time}
P.~Mazur and T.J. Siskens.
\newblock Time correlation functions in the a-cyclic $xy$ model. i.
\newblock {\em Physica}, 69(1):259--272, 1973.

\bibitem{siskens1974time}
T.J. Siskens and P.~Mazur.
\newblock Time-correlation functions in the a-cyclic $xy$ model. ii.
\newblock {\em Physica}, 71(3):560--578, 1974.

\bibitem{fel1998regular}
E.~B. Fel'dman, R.~Bruschweiler, and R.~R. Ernst.
\newblock From regular to erratic quantum dynamics in long spin 1/2 chains with
  an $xy$ hamiltonian.
\newblock {\em Chemical physics letters}, 294(4-5):297--304, 1998.

\bibitem{fel1999regular}
E.~B. Fel'dman and M.~G. Rudavets.
\newblock Regular and erratic quantum dynamics in spin 1/2 rings with an $xy$
  hamiltonian.
\newblock {\em Chemical physics letters}, 311(6):453--458, 1999.

\bibitem{bruschweiler1997non}
R.~Bruschweiler and R.~R. Ernst.
\newblock Non-ergodic quasi-equilibria in short linear spin 1/2 chains.
\newblock {\em Chemical physics letters}, 264(3-4):393--397, 1997.

\bibitem{Mossel}
J.~{Mossel} and J.-S. {Caux}.
\newblock {Relaxation dynamics in the gapped $XXZ$ spin-1/2 chain}.
\newblock {\em New Journal of Physics}, 12(5):055028, May 2010.

\bibitem{lychkovskiy2011entanglement}
O.~Lychkovskiy.
\newblock Entanglement, decoherence and thermal relaxation in exactly solvable
  models.
\newblock In {\em Journal of Physics: Conference Series}, volume 306, page
  012028. IOP Publishing, 2011.

\bibitem{happola2012universality}
J.~H{\"a}pp{\"o}l{\"a}, G.B. Hal{\'a}sz, and A.~Hamma.
\newblock Universality and robustness of revivals in the transverse field $xy$
  model.
\newblock {\em Physical Review A}, 85(3):032114, 2012.

\bibitem{Zhu2010XX}
Z.~Zhu, A.~Aharony, O.~Entin-Wohlman, and P.~C.~E. Stamp.
\newblock Pure phase decoherence in a ring geometry.
\newblock {\em Phys. Rev. A}, 81:062127, Jun 2010.

\bibitem{benderskii2011propagating}
Viktor~Adol'fovich Benderskii and Efim~Iosifovich Kats.
\newblock Propagating vibrational excitations in molecular chains.
\newblock {\em JETP letters}, 94(6):459--464, 2011.

\bibitem{benderskii2013propagation}
VA~Benderskii and EI~Kats.
\newblock Propagation of excitation in long 1d chains: Transition from regular
  quantum dynamics to stochastic dynamics.
\newblock {\em Journal of Experimental and Theoretical Physics}, 116(1):1--14,
  2013.

\bibitem{benderskii2013revivals}
VA~Benderskii, EI~Kats, and AS~Kotkin.
\newblock Revivals in caldeira-leggett hamiltonian dynamics.
\newblock {\em Physics Letters A}, 2013.

\bibitem{niemeijer1967some}
T.~Niemeijer.
\newblock Some exact calculations on a chain of spins.
\newblock {\em Physica}, 36(3):377--419, 1967.

\bibitem{lieb1972finite}
E.H. Lieb and D.W. Robinson.
\newblock The finite group velocity of quantum spin systems.
\newblock {\em Communications in Mathematical Physics}, 28(3):251--257, 1972.

\bibitem{Banchi2010}
L.~Banchi, T.~J.~G. Apollaro, A.~Cuccoli, R.~Vaia, and P.~Verrucchi.
\newblock Optimal dynamics for quantum-state and entanglement transfer through
  homogeneous quantum systems.
\newblock {\em Phys. Rev. A}, 82:052321, Nov 2010.

\bibitem{Banchi2011}
Abolfazl Bayat, Leonardo Banchi, Sougato Bose, and Paola Verrucchi.
\newblock Initializing an unmodulated spin chain to operate as a high-quality
  quantum data bus.
\newblock {\em Phys. Rev. A}, 83:062328, Jun 2011.

\bibitem{haake2010quantum}
F.~Haake.
\newblock {\em Quantum signatures of chaos}, volume~54.
\newblock Springer Verlag, 2010.

\bibitem{Feigenbaum}
M.~Feigenbaum.
\newblock Universal behavior in nonlinear systems.
\newblock {\em Los Alamos Science}, 1(1):4, 1980.

\bibitem{peres1993quantum}
A.~Peres.
\newblock {\em Quantum theory: concepts and methods}, volume~57.
\newblock Kluwer Academic Publishers, 1993.

\bibitem{LL-V}
L.D. Landau and E.M. Lifshits.
\newblock {\em Statistical physics, part 1}, volume~5.
\newblock Pergamon, 1980.

\bibitem{rigol2007relaxation}
M.~Rigol, V.~Dunjko, V.~Yurovsky, and M.~Olshanii.
\newblock Relaxation in a completely integrable many-body quantum system: An ab
  initio study of the dynamics of the highly excited states of 1d lattice
  hard-core bosons.
\newblock {\em Physical Review Letters}, 98(5):050405, 2007.

\end{thebibliography}


\end{document}